\documentclass[english,aps,twocolumn,prX,showpacs,superscriptaddress]{revtex4-2}

\usepackage{graphicx}
\usepackage[caption = false]{subfig}
\usepackage{dcolumn}
\usepackage{bm}
\usepackage{babel}
\usepackage[T1]{fontenc}
\usepackage[latin9]{inputenc}
\usepackage{amsmath}
\usepackage{amssymb}
\usepackage{amsfonts}
\usepackage{booktabs}
\usepackage{microtype}
\DeclareUnicodeCharacter{2212}{-} 
\usepackage{hyperref}
\usepackage{siunitx}
\setcitestyle{numbers,square}
\usepackage{color}
\hypersetup{
	colorlinks = true,
	pdftitle = {},
	pdfauthor = {},
	pdfkeywords = {},
	linkcolor = blue,
	citecolor = blue,
	filecolor = black,
	urlcolor = magenta,
}

\begin{document}
\title{Stability of and conduction in single-walled Si$_{2}$BN nanotubes}

\author{Deobrat Singh$^\S$}
\email{deobrat.singh@physics.uu.se}
\thanks{$^\S$Equal contribution}
\affiliation{Condensed Matter Theory Group, Materials Theory Division, Department of Physics and Astronomy, Uppsala University, Box 516, 75120 Uppsala, Sweden}

\author{Vivekanand Shukla$^\S$}
\email{vivekanand.shukla@chalmers.se}
\thanks{$^\S$Equal contribution}
\affiliation{Department of Microtechnology and Nanoscience-MC2, Chalmers University of Technology, SE-41296, Gothenburg, Sweden}
\author{Nabil Khossossi}
\affiliation{Condensed Matter Theory Group, Materials Theory Division, Department of Physics and Astronomy, Uppsala University, Box 516, 75120 Uppsala, Sweden}
\author{Per Hyldgaard }
\affiliation{Department of Microtechnology and Nanoscience-MC2, Chalmers University of Technology, SE-41296, Gothenburg, Sweden}
\author{Rajeev Ahuja }
\affiliation{Condensed Matter Theory Group, Materials Theory Division, Department of Physics and Astronomy, Uppsala University, Box 516, 75120 Uppsala, Sweden}
\affiliation{Department of Physics, Indian Institute of Technology Ropar, Rupnagar 140001, Punjab, India}
\date{\today}


\begin{abstract}
We explore the possibility and potential benefit of rolling a Si$_{2}$BN sheet into single-walled nanotubes (NTs). Using density functional theory (DFT), we consider both structural stability and the impact on the nature of chemical bonding and conduction. The structure is similar to carbon NTs and hexagonal boron-nitride (hBN) NTs and we consider 
both armchair and zigzag Si$_{2}$BN configurations with varying diameters. The stability of these Si$_{2}$BN NTs is 
confirmed by first-principles molecular dynamics calculations,  by an exothermal formation, an absence of imaginary modes 
in the phonon spectra. Also, we find the nature of conduction varies semiconducting, from semi-metallic to metallic, 
reflecting differences in armchair/zigzag-type structures,  curvature effects, and the effect of quantum confinement. 
We present the detailed characterization of how these properties lead to differences in both the bonding nature and electronic structures.
\end{abstract}

\maketitle


\section{Introduction}\label{sec:intro}

Low dimensional material research has produced exciting results by combining computational predictions, experimental synthesis, and characterization \cite{cheng2021recent}. The immense interest in low dimensional materials is fueled by alluring properties and a broad range of potential applications, such as quantum computing, batteries, electrocatalysis, photovoltaics, electronics, bio-medicals, and photonics \cite{singh2020harnessing,shukla2021electronic, Shukla1275373,xia2014two,singh2020optical,umrao2019anticarcinogenic,singh2020carbon}. There exists a broad range of two-dimensional (2D), one-dimensional (1D), and even dot-like structures. These new materials have 
unique properties and functionalities that are hard to achieve in their three-dimensional counterparts \cite{jain2016computational,marzari2021electronic}.

Computer-assisted materials design today supplements and strengthens the traditional Edisonian-laboratory type exploration. The latter rely on constructive feedback between synthesis and characterization in a trial and error process, and it must, of course, eventually be pursued as we seek progress
on functionality. However, the use of first-principle density functional theory (DFT) allows us to first make reliable predictions of properties, ahead of synthesis. We can therefore focus the more labor-intensive wet-lab activity on the the most promising materials.

There is also good reason to use predictive theory to first screen for likely strong performers and to later explore the result database for beneficial material transformations (into lower-dimensional forms). It is, for example, possible to roll up a 2D sheet into a single-walled nanotube (NT), for example, graphene (or hexagonal boron-nitride) into carbon (or hBN) NTs. Excitingly, theory results for the 2D form may here provide insight on properties of the NT form \cite{Mintmire98}, and give us ideas to rig this transformations to control the resulting NT electronic structure. 
There are a number of material-theory predictions that were later successfully realized in experiments giving trust
in the use of a materials-prediction database. Similarly, the idea of mapping for structure-transformation benefits is today considered a standard option for
a further tuning of the structural and electronic 
properties \cite{cheng2019two,fan2021biphenylene}. 

The Si$_{2}$BN is a particular 2D material 
that is currently attracting significant attention due to its potential use 
in batteries \cite{shukla2017curious}. The theoretical prediction of 
Si$_{2}$BN nanosheet displayed graphene-like planar structure \cite{andriotis2016prediction,sandoval2016stability,fthenakis2021high}, 
having Si-Si-B-N arrangement similar to graphene. 
Si$_{2}$BN sheets show high formation energy similar to graphene and h-BN monolayers, dynamic stability, as well as stability at elavated temperatures. 

\begin{figure*}[htp!]
	    \centering
	    \includegraphics[width=0.8\linewidth]{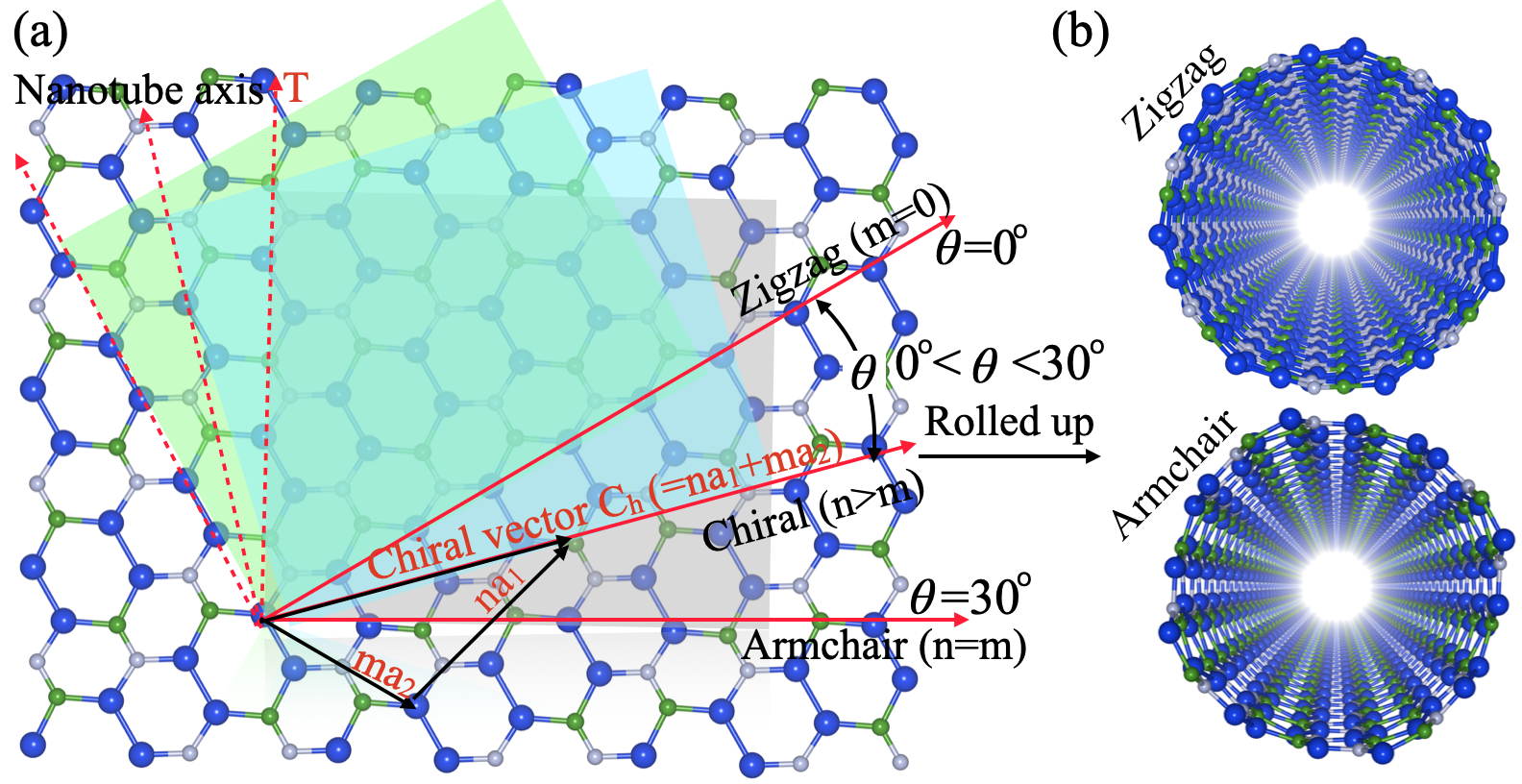}
	    \caption{(a) Schematic representation of rolling a distorted hexagonal Si$_{2}$BN nanosheet with different chiral vector to form single walled Si$_{2}$BN NTs (b) Schematic illustration of of top and side views of zigzag (7,0) and armchair (4,4) NT geometries with $\sim$1.5 nm diameter.The blue, green and light gray color represent the Si, B and N atoms respectively.}
	    \label{arm-zig}
\end{figure*}

It is natural to next inquire whether there is also here potential usefulness in a Si$_{2}$BN sheet
rolling into a 1D NT form?

In this paper, we propose to explore potential benefits of such Si$_{2}$BN transformations. To that end, we map the structure and electronic-structure impact of various forms of sheet rolings. A key motivating factor is that Si$_{2}$BN systems presents us with a richer chemistry and therefore also a richer physics of the corresponding 1D form, compared to graphene and carbon NTs. The Si atom has Si, B, and N as nearest neighbours, while each B (N) has two Si atoms and  one N (B) atom as nearest neighbours. The Si atoms finds themselves in electron-deficient position
and the variation in bonds works as an electron reservoir for an adatom or a molecule on the sheet (or NT) surface. This characteristics of 2D Si$_{2}$BN has been successfully explored 
in a variety of applications such as hydrogen 
storage \cite{singh2017high,hu2020si}, metal-ion 
batteries \cite{shukla2017curious}, gas 
sensing \cite{hussain2019efficient}, active 
catalyst for hydrogen evolution reactions \cite{singh2019emergence} in the short 
span of time that has passed since their first synthesis. Doping might also affects the weakly metallic electronic structure in the search for a high-carrier mobility or for thermoelectric applications.\cite{mahida2021hydrogenation,fthenakis2019structural,singh2018achieving,mahida2019influence} A controlled rolling into (single-walled) Si$_{2}$BN NTs may allow us to also realize more of a semiconducting nature conduction, as desired for
various potential applications \cite{andriotis2016prediction,fthenakis2019structural,singh2018achieving,mahida2019influence}.

Also, the Si$_{2}$BN presents intriguing possibilities for using the NT formation to 
control the nature of conduction even beyond what is possible in the carbon and hBN NT counterparts.\cite{Mintmire98} Interestingly, the hexagons in Si$_{2}$BN planer structures are distorted in shape (unlike graphene and silicene) owing to variations in the electronegativity and in the covalent radii. This, in turn, implies the presence of strain, although we
are still staying close to the sp$^{2}$-hybridized chemical bonding that characterize graphene.

Mintmire \textit{et al.} showed that the orientation
of the rolling (NT) axis (relative to the 
graphene hexagons) determines whether
we get a metallic or semiconducting carbon NT, in
a universal relationship \cite{Mintmire98} 
(that applies without pronounced rehybridization).
The Si$_{2}$BN sheet does not show a Dirac-cone behavior at the Fermi level, but the resulting NTs still retain a sharp density of state around the Fermi level. \cite{shukla2017curious, andriotis2016prediction} On the one hand, we can therefore expect the general ideas to guide us.\cite{Mintmire98} On the other hand, unlike for carbon NTs, we can now also adjust the extent that
we let the Si-atom site carry the 
main load
of rolling-induced strain. This fact provides us with additional options to control the electronic structure details in the Si$_{2}$BN NTs.

We use first-principles DFT calculations and transport studies to demonstrate that such Si$_{2}$BN NT formation is indeed possible (in terms of structural stability) and allows control
of the nature of conduction. In fact, motivated by the potential for a richer band gab engineering, we systematically investigate the electronic properties and bonding characteristics of a range of single-walled Si$_{2}$BN NTs of various sizes and choice of rolling axis. For two main options for Si$_{2}$BN  rolling (corresponding to armchir and zigzag rolling of carbon NTs) we validate the mechanical stability by documenting an absence of vibrational instabilities and by tracking the systems using Ab Initio molecular dynamics. More broadly we use a crystal-orbital Hamiltonian population (COHP) analysis to document that the bonding between the Si-Si, Sb-B/N, and B-N atoms retains a pronounced covalent nature; A special case
is discussed in the supplementary materials (SM).
We take the COHP characterization as an indicator that 
the new NTs are stable. We also take this information as a sign that the Si$_2$BN NTs will generally resist (as do carbon NTs) forming strong chemical bonds in connection with functionalization and doping. We generally expect that the interaction with the environment will occur through weak chemisorption \cite{Dion,behy13,bearcoleluscthhy14,Berland_2015:van_waals,Thonhauser_2015:spin_signature,MehMuMur18,hyldgaard2020screening}.

This paper is organized as follows. The following section \ref{method} describes the theoretical approaches. The results and discussion section III  is separated into three main themes. Subsection \ref{stru-stability} reports our predictions of the resulting NT structures and discusses the mechanical and
chemical stability, subsection \ref{bonding-info} 
reports our exploration of the electronic properties, while
subsection \ref{bonding-info}, presents and discusses the results of our chemical-bond analysis. We conclude with a summary and outlook section \ref{conclusion}.



\section{Computational method}
\label{method}

First-principles calculations are employed with the plane-wave basis projector augmented wave (PAW) method in the framework of density-functional theory (DFT) \cite{hohenberg1964inhomogeneous} coded in Vienna \textit{Ab-Initio} Simulation Package (VASP) software \cite{kresse1996efficient}. For the exchange-correlation potential, we use the generalized gradient approximation (GGA) in the form of the Perdew-Burke-Ernzerhof (PBE) functional. The energetics of rolling of a sheet ribbon into a NT involves balancing a large exothermal process of sealing the ribbon edges against significant cost of mechanical deformations. We ignore  long-range dispersion effects, for example, as described in the vdW-DF method \cite{frostenson2021hard,hyldgaard2020screening} since we expect that both of  these competing energies are primarily defined by the presence of strong covalent bonds in the Si$_2$BN material.

A plane-wave basis-set energy cutoff value is first converged and then taken to be 500 eV throughout or calculations, including first-principle molecular dynamics simulations. The total energy was minimized until the energy variation in successive steps became less than 10$^{-6}$ eV in the structural relaxation and the convergence criterion for the Hellmann-Feynman forces was taken to be 10$^{-3}$ eV/\AA. We utilized the 24x1x1 $\Gamma$-centred k-point sampling for the unit cell in single-walled Si$_2$BN NTs for the electronic structure. The density of states (DOS) was plotted with 0.026 eV Gaussian broadening (to allow us to easily
interpret predictions for the room-temperature nature of conduction). We studied the NT in unit cells and supercells in which we have added a 15 \AA\ vacuum spacing in the lateral dimensions to minimize the spurious interaction between the periodically repeated NT images. 

The charge transfers in different elements in the NTs structures were determined by the post-processing tool Bader analysis \cite{henkelman2006fast} implemented in VASP. The first-principles molecular dynamics (FPMD) was applied to check the thermal stability at high temperatures. The canonical (fixed number of atoms, N, fixed volume, V, and a fixed temperature, T) NVT  ensemble was used during FPMD simulation for 5 ps with the time steps of 2 femtoseconds (fs) and the temperature was controlled via Nos\'e-Hoover method \cite{martyna1992nose}. We used the PHONOPY code to calculate the phonon dispersion spectra computed via density functional perturbation theory (DFPT) \cite{mathew2014implicit}. 
We calculated the crystal orbital Hamiltonian population (COHP) using the e Local Orbital Basis Suite Towards Electronic-Structure Reconstruction (LOBSTER) software package \cite{maintz2016lobster}.

 \begin{figure}[htp!]
	    \centering
	    \includegraphics[width=1.0\linewidth]{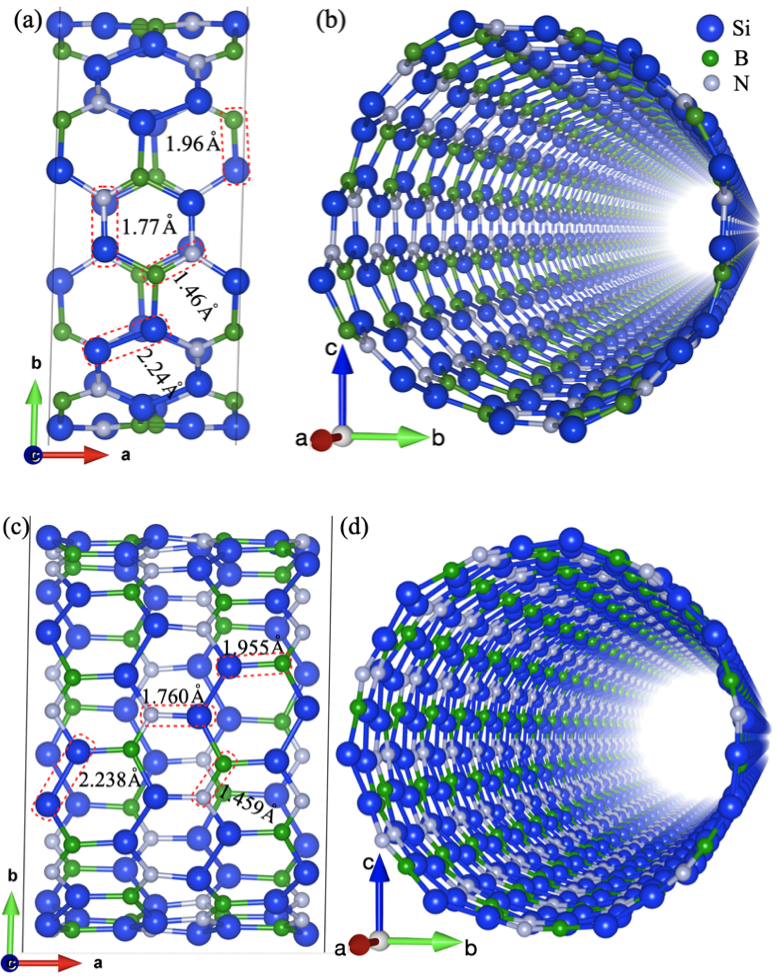}
	    \caption{Optimized structures of (a,b) armchair (4,4) and (c,d) zigzag (7,0)  Si$_{2}$BN  nanotube with top view and other orientation. The blue, green and light gray color represent the Si, B and N atoms respectively. The bond lengths in fully relaxed ground state structures are indicated in respective figures.}
	    \label{F1}
    \end{figure}

Finally, the electronic transport calculations were performed using the non-equilibrium Green function (NEGF) formalism, ignoring inelastic and dephasing scattering events. Here we used the quantum transport code TranSIESTA that is implemented along with SIESTA \cite{papior2017improvements,soler2002siesta}. \par

\section{Results and discussion}
\label{ResultsDiscusion}

\subsection{Structural characterization and stability} \label{stru-stability}

The single-walled Si$_{2}$BN NTs are formed from a ribbon of graphene-like Si$_{2}$BN monolayer. The monolayer ribbon is rolled up into a hollow seamless cylinder in a similar way as carbon NTs \cite{Mintmire98} and Type-I BC$_{2}$N NTs can be rolled \cite{miyamoto1994chiral}. To provide a 
simple nomenclature, we use $\mathbf{a}_{1,2}$ to denote the unit vectors for the hexagonal structure in the sheet and assume that the
NT arises as a mathematical construction. First starting with any atom, we consider
a so-called chiral vector 
\begin{equation}
    C_{h} = n\mathbf{a}_1 + m\mathbf{a}_2 \, ,
    \label{eq:chiralVector}
\end{equation}
that will take us to another equivalent atom
in the sheet (taking $m$ and $n$ directions
of the basis vectors). We next assume that we make a ribbon by cutting the sheet at right angles at the start and just before the very end of this vector; These edges define the
resulting NT axis. Finally, we see the NT formation as resulting after bending the ribbon and rebonding at the edges. Equation (\ref{eq:chiralVector}) is also called the roll up vector since it connects repeated ribbon images of the sheets and therefore defines
the atomic structure in the circumference of the resulting cylinder.

Figure \ref{arm-zig}(a) identifies the chiral angle
$\theta$ that the roll-up or chiral vector
has with the hexagonal-sheet basis vectors. The panel also illustrates two main cases, `Zigzag' NTs that corresponds to $\theta =0$ and `Armchair' NTs that corresponds to $\theta = 30$. For these NT constructions one must
use ribbons with chiral vector set by $(n,0)$
and $(n, m=n)$ in-sheet steps, respectively.
Fig. \ref{arm-zig}(b) shows the schematic of resulting zigzag and armchair Si$_{2}$BN NTs that both approximately have a 1.5 nm diameter.

Importantly, Fig.\ 1(a) also clarifies that Si$_{2}$BN have fever possibilities for rolling into a NT (compared with graphene). This is because the sheet is a triple-element system
and we we must have periodicity as we revolve
around the cylinder. A chiral vector cannot start with a Si atom and end with a B or N atom. Nevertheless we have constructed and characterized a total of 5 zigzag NTs and 7 armchair NTs of various diameters.





Figure \ref{F1}(a,b) shows our computed results for the fully relaxed structures of two moderately large Si$_{2}$BN NTs. We contrast a zigzag and an armchair form, top and bottom set of panels, respectively. In both cases, we aimed to have approximately a 1.5 nm diameter, and therefore considered the armchair (4,4) and zigzag (7,0) NTs for detailed stability analysis, below. 

The zigzag and armchair Si$_{2}$BN NTs have 64 (32-Si, 16-B, and 16-N) and 112 (56-Si, 28-B, and 28-N) atoms in the unit cell, respectively. The unit cell lengths along the axial directions are 6.45 \AA{} and 11.06 \AA{} for the armchair and zigzag directions. 
We use an initial structure guess to start a DFT study of the atomic relaxation in both 
NT case, and find that the curvatures influence the bond lengths slightly. The fully relaxed NT bond lengths are Si-Si (2.24 \AA{}), Si-B (1.96 \AA{}), Si-N (1.77 \AA{}), and B-N (1.46 \AA{}), that is, very similar to those that characterize the 2D Si$_2$BN monolayer form (2.246, 1.951, 1.756, and 1.466 \AA{}) \cite{singh2019emergence}. 

To further probe the energetic stability of the fully relaxed NTs, we calculated the cohesive energies and compared them with various other NTs. Our NT cohesive energy results may depend weakly on the choice of pseudopotentials that we use for the DFT calculations. We compare with the available literature values for a range of other NTs, Table \ref{coh}.

\begin{table}[ht!]
\begin{ruledtabular}
\begin{center}
	\small
	\caption{Cohesive energy (in eV/atoms) of single walled Si$_{2}$BN NTs in armchair (4,4) and zigzag (7,0) forms. Atomic reference energies for Si, B and N are calculated using spin polarised calculations (using unit-cell sizes with 15 \AA \ vacuum spacing between the repeated images of the NTs). These cohesive energies of Si$_{2}$BN  nanotubes are compared with available cohesive energies for various other nanotubes in the literature.}
	\label{t1a}
	\begin{tabular}{c c c}
		System & $E_{coh}$& Ref.\\
		\hline
		Borophene & 3.44 &\cite{fazilaty2021investigating}\\
		Silicene & 4.014 &\cite{zhang2017chirality}\\
		Blue phosphorene & 4.417 &\cite{bhuvaneswari2021molecular}\\
		Graphene & 7.906, 8.72 &\cite{shin2014cohesion,girifalco2000carbon,juarez2017stability}\\
		Graphyne & 7.02 &\cite{kang2015electronic}\\
		Boron nitride & 7.59 & \cite{juarez2017stability}\\
		Si$_{2}$BN nanotube Zigzag/Armchair & 4.92/4.90 &This work\\
	\end{tabular}
    \label{coh}
    \end{center}
    \end{ruledtabular}
\end{table}

We find that that armchair and zigzag Si$_2$BN NTs cohesive energies (4.92 eV and 4.90 eV,
respectively) fall in the middle 
of those that applies for other types of NTs. Our results suggest
that the Si$_2$BN NTs are stable compared to silicon, boron, and phosphorus-based nanotubes.
There is a lower stability in comparison to values that are also experimentally tested for the C NTs and hBN NTs. However, we still expect metastability because 
(as we document in the following
subsection) all bonds retains a pronounced covalent character, implying that there are dramatic
costs in breaking any of the Si$_2$BN bonds.

To further confirm the energetic stability of Si$_2$BN NTs, we also
investigate the impact of increasing the NT diameters to $\sim$2.5 nm, in the armchair (7,7) and the zigzag (12,0) NTs. We find that the cohesive energy remains unchanged (4.89 eV and -4.91 eV/atom for the larger armchair and zigzag Si$_2$BN NTs, respectively).
Additionally, we characterize the cohesion and stability of the smallest possible diameter NTs, 
performing both DFT relaxation studies and a complete phonon characterization for the armchair (1,1) and zigzag (2,0) NTs. A detailed description of those test can be found in the SM. There is eventually a dynamical instability, but these small-NT studies also help document that the Si$_2$BN NTs retain mechanical stability over a broad range of diameters.

Table \ref{strain} tracks both the cohesive-energy and the strain energy for both the armchair and zigzag NTs across the varying diameters. The strain energy is calculated by subtracting the nanotube energy per atom with the per-atom energy of Si$_2$BN in planner form. 

We find that the strain energy decreases with the diameter, and as expected, a higher stability of larger NTs. A decreasing impact of strain also manifests itself when we track the length of Si-Si, Si-B, Si-N, and B-N bonds in the NTs. Table \ref{bond} clearly shows that bonds are strained in low diameter tubes, and this strain does eventually cause a dynamical instability, see SM. 

\begin{figure}[htp!]
	    \centering
	    \includegraphics[width=0.99\linewidth]{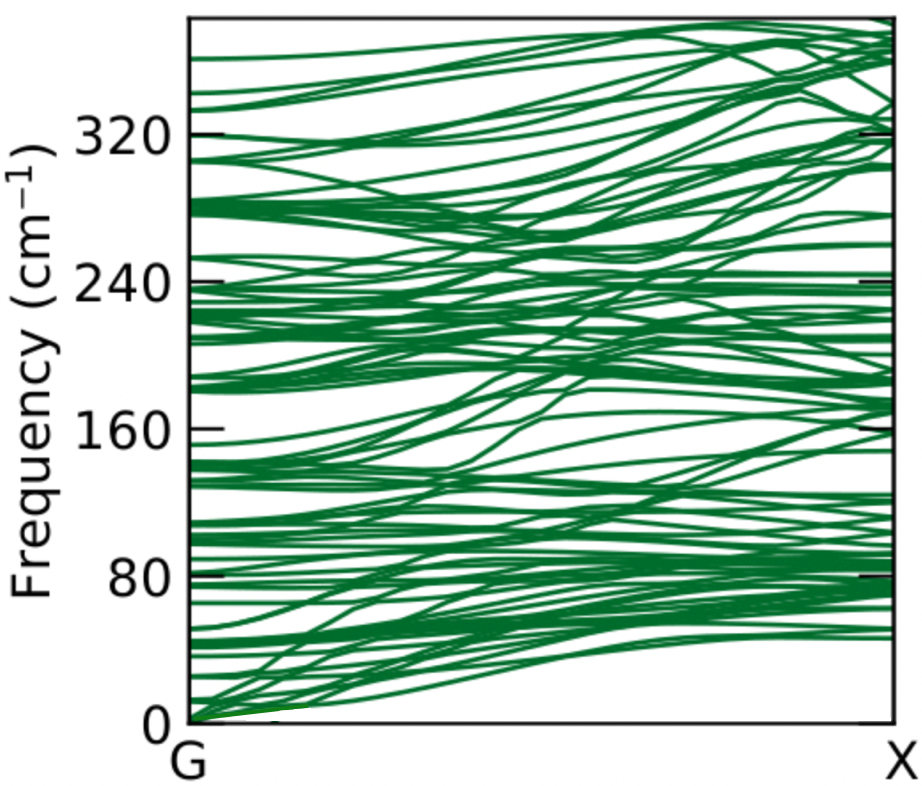}
	    \caption{Phonon dispersion profile of armchair Si$_{2}$BN (4,4) nanotube. Absence of imaginary frequencies in phonon dos show the dynamic stability of the nanotube.}
	    \label{F2}
    \end{figure}

Figure \ref{F2} shows the phonon dispersion band structure for the armchair (4,4) nanotube  that has a $\sim$1.5 nm diameter. We find no imaginary phonon modes in the spectra, thus documenting also  dynamical stability of this NT. A similar conclusion holds for the zigzag NTs of similar diameters.

The phonon characterizations also allow us to discuss thermodynamic properties. We focus on the free energy, the specific heat at constant volume C$_{\rm V}$ \cite{xiao2003specific,einollahzadeh2016studying,madsen2006boltztrap}, and the vibrational entropy. Figure S2 and of the SM document and present our DFT-based predictions for those properties as a function of the temperature.  

\begin{table*}[ht]
\begin{ruledtabular}
\begin{center}
	\small
	\caption{Comparison of chiral vector (n,m), diameters in nanometer (nm) and the number of atoms in the unit-cell during calculation for various Si$_{2}$BN armchair (ac) and zigzag (zz) NTs. 
	We also list computed values for the strain energy $E_{\rm strain}$ and the nature of conduction and compares the semiconductor bandgap $E_g$ (when present).}
	\label{t1b}
	\begin{tabular}{c c c c c}
		acSi$_{2}$BN nanotube & No. of atoms & NT diameter (nm) & E$_{\rm strain}$ (eV/atom) & E$_g$ (eV)\\
		\hline
		(1,1) & 16 &$\sim${0.4} & 0.184 & Metallic\\ 
        (2,2) & 32 &{0.7} & 0.086      & 0.09 (Indirect)\\ 
        (3,3) & 48 &{1.06} & 0.050     & Semi metallic\\
        (4,4) & 64 &{1.5} & 0.033     & Metallic\\ 
        (7,7) & 112 &{2.5} & 0.045     & Metallic\\
        \hline
		zzSi$_{2}$BN nanotube & No. of atoms & diameter (nm) & E$_{strain}$ (eV/atom) & E$_g$ (eV) \\
		\hline
		(2,0) & 32 &$\sim${0.4} & 0.117 & 0.11 (Direct)   \\
        (3,0) & 48 &{0.61} &      0.078 & 0.14 (Indirect)  \\
        (4,0) & 64 &{0.82} &      0.066 & 0.05 (Indirect)   \\
        (5,0) & 80 &{1.03} &      0.054 & 0.18 (Indirect)   \\
        (6,0) & 96 &{1.23} &      0.041 & Semimetallic      \\
        (7,0) & 112 &{1.5} &      0.032 & Semimetallic      \\
        (12,0) & 192 &{2.5} &     0.026 & Metallic           \\
	\end{tabular}
    \label{strain}
    \end{center}
    \end{ruledtabular}
\end{table*}

    \begin{figure}[h!]
	    \centering
	    \includegraphics[width=0.99\linewidth]{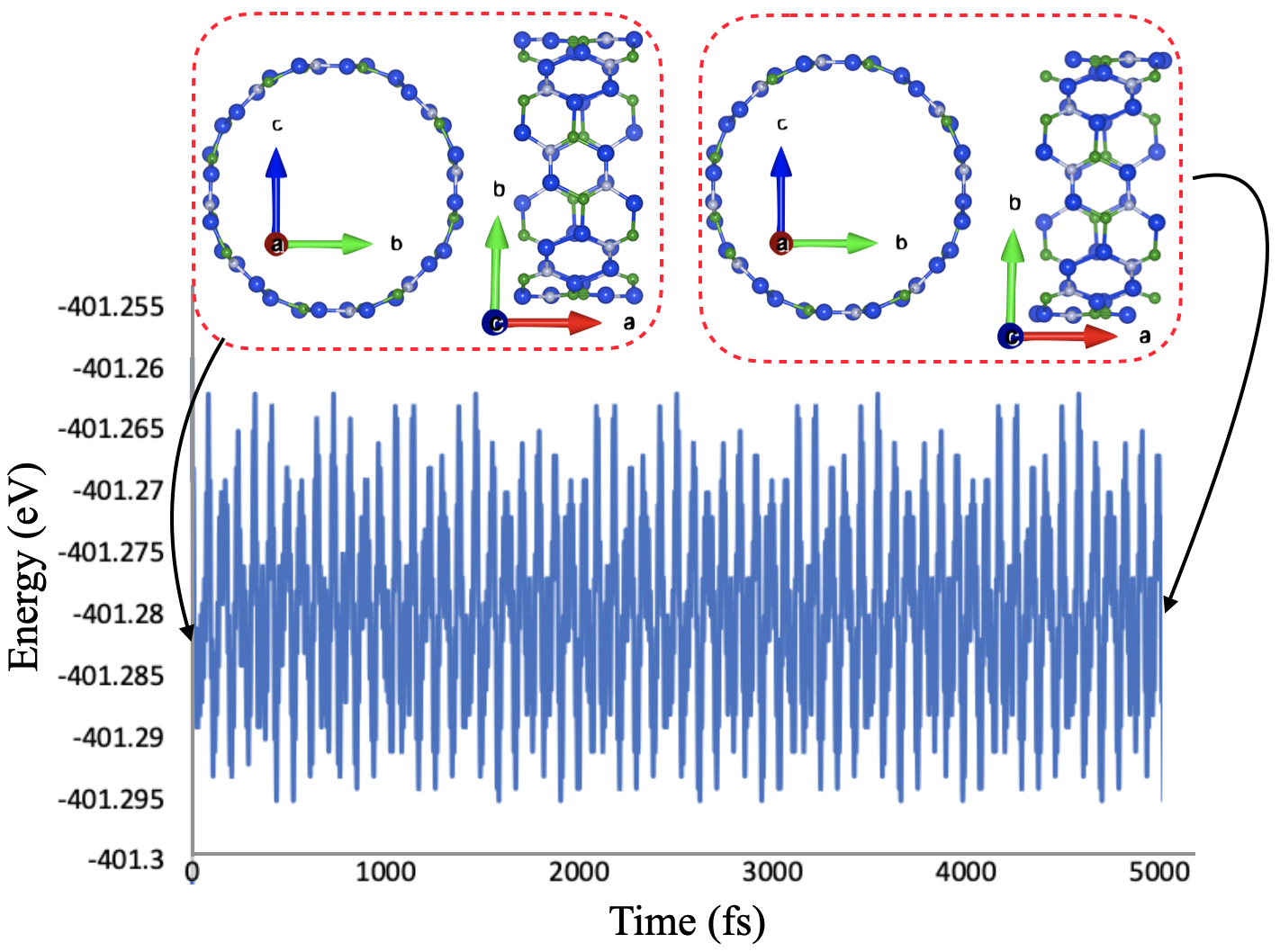}
	    \caption{Variations of the total energy for armchair (4,4) Si$_{2}$BN nanotube during First principle molecular dynamics simulation at higher temperature 1000 K for 5 ps. The small variation in total energies documents that this nanotubes is stable also at elevated temperatures.The inserts above the energy variations shows the initial and final structure during the molecular dynamics simulations with similar bond length and structure.}
	    \label{FPMD}
    \end{figure}
    
We find that the (armchair) NT free energy decreases monotonously with increasing temperature, whereas vibrational entropy tends to increase. The heat capacity C$_{\rm V}$ increases very rapidly at low temperatures, reflecting an expected compliance with the Debye law (T$^{3}$) \cite{gusev2019quasi}, At temperatures above 600 K, the C$_{\rm V}$ rolls over to a constant, hence also complying with the Dulong-Petit law \cite{dulong1819recherches}. 

Our access to the predicted phonon
density of states, Fig.\ S2 of the SM, also allows us to provide a simple discussion of the
vibrational contribution to
the low-dimensional thermal conductivity \cite{HyGDM1,Chen96,HyGDM2,Chen97,Balandin98,Chen98,ElHy05,ElHy06,GDMreview}. The lattice part of the NT thermal conductivity is itself proportional to the heat capacity C$_{\rm V}$ \cite{singh2021dimensionality} and to the expected mean-free phonon path. The latter reflect a `wire' scattering time \cite{HyGDM1,ElHy06} and the relevant (along-axis) phonon group velocity that involves a projection
of the density of state information \cite{HyGDM2,Balandin98,ErHyLin} (along the
NT axis). For low-dimensional
systems it is not, in general, 
easy to extract this information from density of state but we can
proceed as follows. 

First, we observe that the heat capacity is significantly higher for the Si$_2$BN NT than in  C NT s \cite{xiao2003specific,PhysRevB.67.045413,zhang2003specific}. This itself suggest a high lattice thermal conductivity. 

Next, we consider a DFT-based evaluation of the bonding stiffness
for the Si$_2$BN NT. We provide a spring-constant analysis from our calculations of the harmonic-force-constant tensor that describe coupling between nearest-neighbor atom pairs \cite{lee2014resonant,gewahy20}. The spring constants, in this case, are 12.57, 17.15, 13.16, and 26.54 eV/\AA{} for atom pairs Si-Si, Si-N, Si-B, and B-N, respectively. These values are higher than the values that applies for most of the materials that have previously been explored \cite{sajjad2019ultralow,gu2014phonon,sajjad2021impact}. The large Si$_2$BN NT spring-constant values (reflecting stiff bonds) 
explains the large predicted value for the high-temperature Dulong-Petit C$_{\rm V}$ limit (and equivalently, a large acoustic-cutoff frequency $\omega_{\rm cut}= 360$ cm$^{-1}$, indicated by a dotted line in Fig. S2 and by the plotting range in Fig.\ \ref{F2}). The large C$_{\rm V}$ value will itself impact the Si$_2$BN NT thermal conductivity. 

Meanwhile, the NT phonon group velocity is set by the ratio of the string constants and the masses, see for example, Ref.\ \onlinecite{GDMreview,HyGDM2}. The weighted impact of such phonon group values should ideally be calculated as a function of the temperature \cite{ErHyLin}. However, since the phonon dispersion is here strictly 1D, the large value of Dulong-Petit limit itself guaranties a large average along-axis group velocity \cite{HyGDM1,ElHy06}. 

Looking at the vibrational DOS, Fig.\ S2 in the SM, we see that the phonons fall in two classes: A lower group, bounded by the vertical dotted line at $\omega_{\rm cut}=381.06$ cm$^{-1}$, and an upper group that can be interpret as optical phonons. The lower group contains phonons can be seen as connecting the acoustic phonons and vibrations at $\omega_{\rm cut}$. We shall refer to the latter as an effective acoustic cutoff. The average group velocity of the transport carrying modes is clearly bounded by $\omega_{\rm cut}/k$, where the inver length $k$=0.0775 is the propagation vector, given by the reciprocal-space distance between X and Gamma. 

While the ratio $\omega_{\rm cut}/k$ is not a quantitative measure of the average group velocity, it suffices
for a comparison of phonon transport in similar NT, for example carbon and Si$_2$BN NTs. That is, we can compare the magnitudes of $\omega_{\rm cut}/k$ to get an indication of whether a NT is a fair heat carriers. We find that the Si$_2$BN value compares with the value that we also here estimate for a (5,5) C NT: 17 km/seconds versus 14.95 km/seconds \cite{maruyama2003molecular}.

In summary,  there are good reasons to expect a significant vibrational contribution to the
Si$_2$BN NT thermal conductivity.

Figure \ref{FPMD} and Figure S3 in SM presents an extra, independent check of the NT stability that we also provide. The figure summarizes the results of an full
first principle molecular dynamics (FPMD) simulations. It shows the structure of the armchair/zigzag NT at the initial and final stage of FPMD simulation performed at 1000 K for 5 ps. The
figure also shows the computed variations that we find om the total energy of the system as a function time steps. We find that this variation is minimal and that there has been no bond breaking during the MD run. The study confirms that we can also expect a degree of thermal stability, even at high temperatures.

We find, in particular, that the dynamics is characterized by the same structure that exists in the initial and final NT configurations. i.e., the
bond lengths does not change. For that analysis, we extracted a 
few snap-shot images 
and then fully relaxed the ionic positions of those configuration with a regular DFT study (until the
forces converged up to 0.01 eV/atom). Among those fully relaxed NT variants, we find no significant differences from the extracted structures. 

Interestingly, we also learned from putting the smallest diameter NTs onto such MD-based stability test. Fig. S4 of the MD shows
that the armchair (2,0) NT has higher fluctuation in total energies with time in comparison to zigzag (1,1) NT. We see that as we approach the very small NTs (that have an actual dynamical instability) there are incipient fluctuations in the MD runs. In other words, the MD simulation test is able to correctly identify unstable NTs, when relevant. We consider our MD testing a robust, independent confirmation of the expected high-temperature stability for the regular-sized Si$_2$BN NTs.

\begin{table}[h!]
\begin{ruledtabular}
\begin{center}
	\small
	\caption{Bond length (\AA{}) between the atoms in fully relaxed ground states of single walled armchair (ac)  and zigzag (zz) Si$_{2}$BN NTs with different chiral vector (n,m) compared with the respective bond lengths in Si$_2$BN 2D sheet. Small-diameter NTs are highly strained in the bond lengths in comparison to large-diameter NTs.}
	\begin{tabular}{c c c c c}
	    System & Si-Si & Si-B & Si-N & B-N \\
	    \hline
	    2D-sheet & 2.246  & 1.951 & 1.756 & 1.466 \\
	    \hline
		AC(1,1) & 2.393 &2.078 & 1.791 & 1.426 \\
        AC(2,2) & 2.292 &2.002 & 1.749 & 1.441\\
        AC(3,3) & 2.248 &1.977 & 1.758 & 1.447\\
        AC(4,4) & 2.240 & 1.96 & 1.77  & 1.460  \\
        AC(7,7) & 2.246 &1.951 & 1.756 & 1.466 \\
        \hline
		ZZ(2,0) & 2.325 & 2.037  & 1.789 & 1.451 \\
        ZZ(3,0) & 2.280 & 2.014  & 1.741 & 1.443 \\
        ZZ(4,0) & 2.261 & 1.996  & 1.746 & 1.438 \\
        ZZ(5,0) & 2.238 & 1.992  & 1.758 & 1.441 \\
        ZZ(6,0) & 2.239 & 1.958  & 1.760 & 1.457   \\
        ZZ(7,0) & 2.238 & 1.955  & 1.760 & 1.459  \\
        ZZ(12,0)& 2.244 & 1.951  & 1.756 & 1.466   \\
	\end{tabular}
    \label{bond}
    \end{center}
    \end{ruledtabular}
\end{table}

\subsection{Chemical bonding information}
\label{bonding-info}
 \begin{figure}[h!]
	    \centering
	    \includegraphics[width=0.99\linewidth]{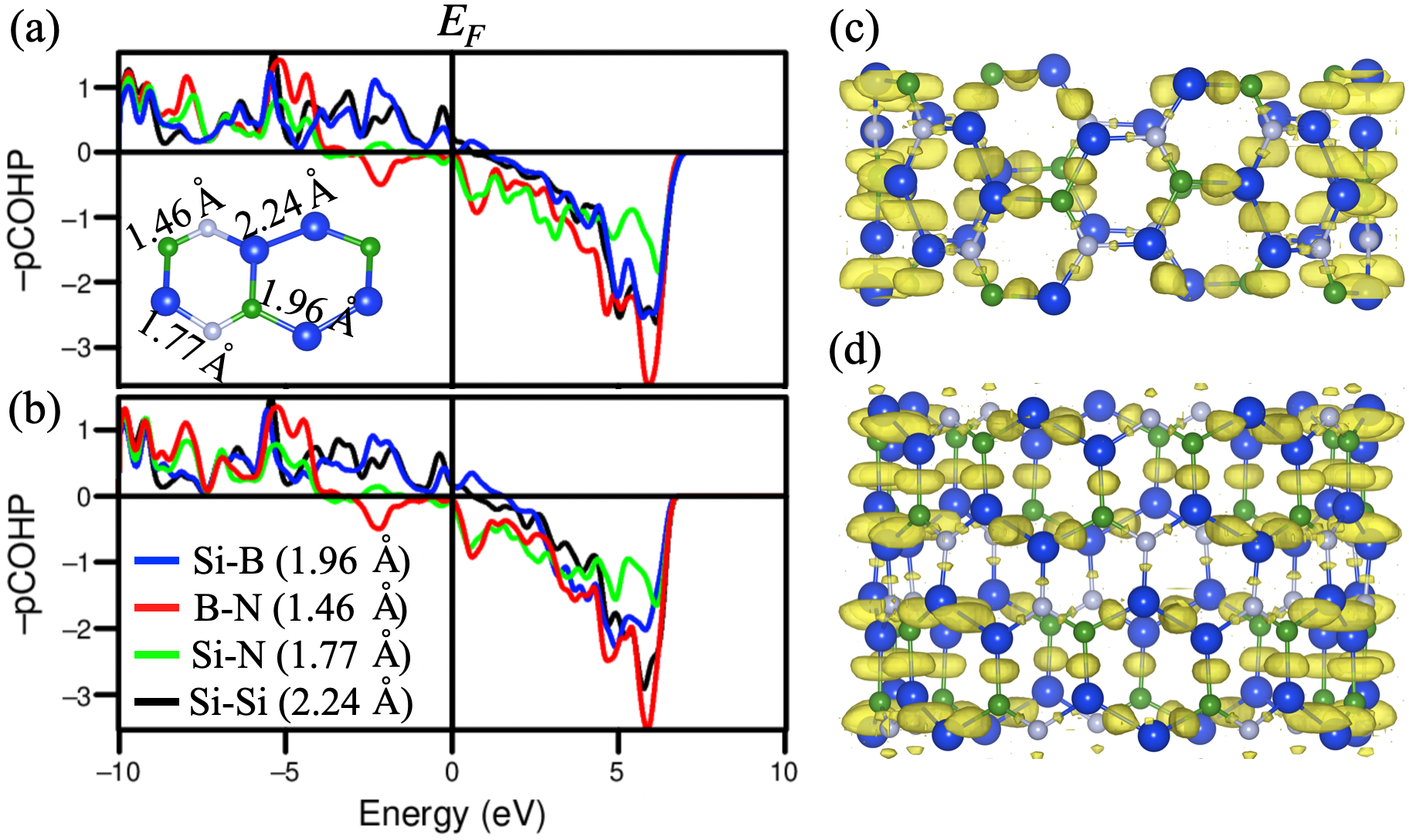}
	    \caption{Projected crystal orbital Hamilton Population (-pCOHP) analysis for (a) armchair (4,4) and (b) zigzag (7,0) Si$_{2}$BN NTs summed up over the pairs Si-Si, B-N, Si-B and Si-N. The zero level shows the Fermi energy ($E_{F}$). The -pCOHP provides information about the contribution of a specific bond (bonding or anti bonding) to the band energy. The positive value on y-axis represent the bonding contribution and negative value on y-axis shows the antibonding contribution. Electron localization function of (c) armchair (4,4) and (d) zigzag (7,0) Si$_{2}$BN NT. }
	    \label{cohp-elf}
    \end{figure}

\begin{figure*}[htp!]
	    \centering
	    \includegraphics[width=0.99\linewidth]{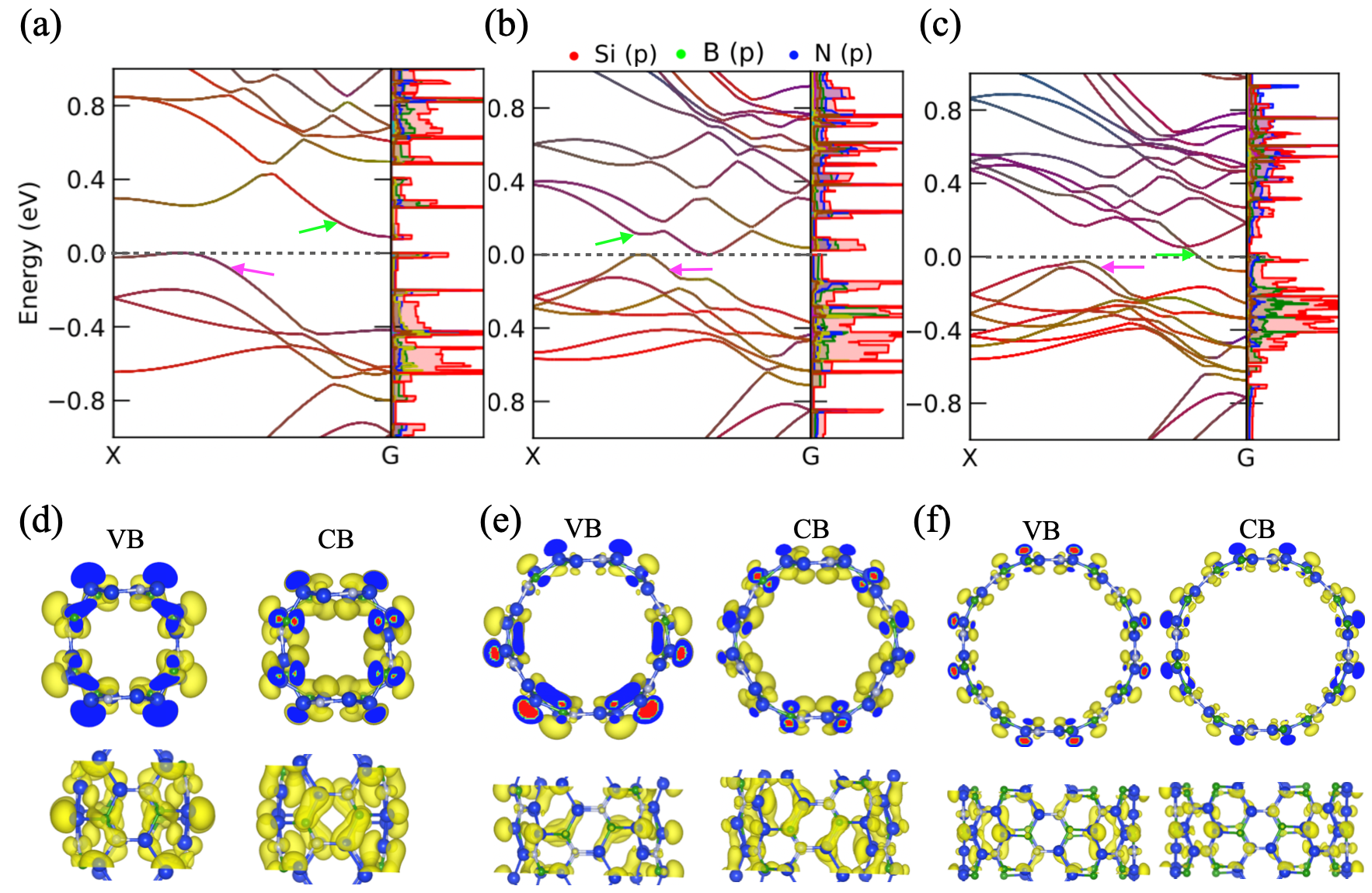}
	    \caption{\textit{Top row:} Orbital- and element-resolved comparison of the electronic band structure and corresponding projected density of states for armchair direction (a) (2,2) (b) (3,3) and (c) (4,4) NTs. We focus on the region
	    near the Fermi level, which is set at zero. Pink and green arrows arrows the highest valence band (VB) and lowest conduction band (CB), respectively. \textit{Bottom row:} Band-resolved mapping of the charge density below the Fermi level and probability density above the Fermi level, with panels (d), (e) and (f) corresponding to the system characterized in panels (a), (b), and (c), respectively. The orbital representation is extracted at the lowest CB (highest VB) energy location of the bandstructure parts identified by green (pink) arrows. They identify the CB/VB orbital nature that defines transport properties in the set of armchair NTs that have increasing diameters.}
	    \label{band-ac}
\end{figure*}

Beyond the two main-interest
(normal sized) NT formation processes, above, we have also sought a broader mapping of stability in the NT formation and the size impact that will exist on properties. Since this exploration must begin with a mechanical stability
analysis, we have sought a simpler tool than using a combination of both phonon calculations and FPMD simulations, 
for every considered Si$_2$BN NT.

The COHP analysis partitions the band-structure variation into contributions that identifies the binding nature, that is, nonbonding, and antibonding contributions. This is done using projections on localized atomic basis sets. We use it as extracted from a projection of our planewaves DFT results, a so-called pCOHP analysis.  We find that the pCOHP analysis is useful for our wider exploration and discussion: It allows us to document 
the extent that the NT bonding retains a pronounced covalent character even after we have imposed significant extra strain in completing the NT rolling.

Figure \ref{cohp-elf}(a,b) tracks the chemical bonding as revealed in a pCOHP  analysis \cite{dronskowski1993crystal,eivari2017two} for the armchair (4,0) and zigzag (7,7) NTs, respectively. The inset shows the bond lengths between the atoms and color combinations for different bonding pairs. The main panels
display an energy-resolved mapping of the nature of the chemical bonding that underpins the delocalized electronic structure \cite{deringer2015chemical}. 
Positive (negative) values reflects bonding (anti-bonding) contributions
as the emerge on orbitals corresponding to different bonds among the constituent atoms \cite{shukla2017curious}.

Figure \ref{cohp-elf}(a,b) shows that all the bond pairs have stable chemical bonding interactions because few anti-bonding states are located at or below the Fermi level. We can see some bonding emerging also above the Fermi level (mainly appearing from Si-Si and Si-B bonds); The presence of such bonding
states implies that the conduction band is able to accumulate also some extra electrons, i.e., reflecting a metallic signature. Anti-bonding states dominates well above the Fermi level, Fig. \ref{cohp-elf}(a,b), but such a character is limited to B-N bond contributions from just below the Fermi level, in both the armchair and zigzag NTs. 


The corresponding integrated-COHP (ICOHP) measure has already been used to measure bond strengths in various materials \cite{khazaei2019novel}. The ICOHP values for Si-B, Si-Si, Si-N and B-N bond-pairs are found to be -5.74 (-5.77), -6.20 (-6.19), -6.28 (-6.25) and -8.75 (-8.69) eV for armchair (zigzag) directions, respectively. The B-N and Si-N come out to be the strongest bond-pairs interaction compared to other bond pairs in the Si$_2$BN NT. However, it should be noted that ICOHP mainly measures the strength of the covalency of a bond, but not its ionicity. 

Finally, Fig.\ \ref{cohp-elf}(c,d) shows a map of the electron localization function (ELF), again revealing the nature of bonding between Si, B, and N atoms. The electron density is mainly localized between the Si-Si, Si-B, Si-N, and B-N atoms, which indicates that the Si-Si, Si-B, Si-N, and B-N atoms form strong covalent bonds. This observation supports the assumption of the possibility of excellent mechanical properties and the high hardness of these NTs \cite{liao2009chemical,naebe2014mechanical}.

\subsection{Electronic Structure and transport properties}
\label{elec-prop}
\begin{figure*}[ht]
	    \centering
	    \includegraphics[width=0.99\linewidth]{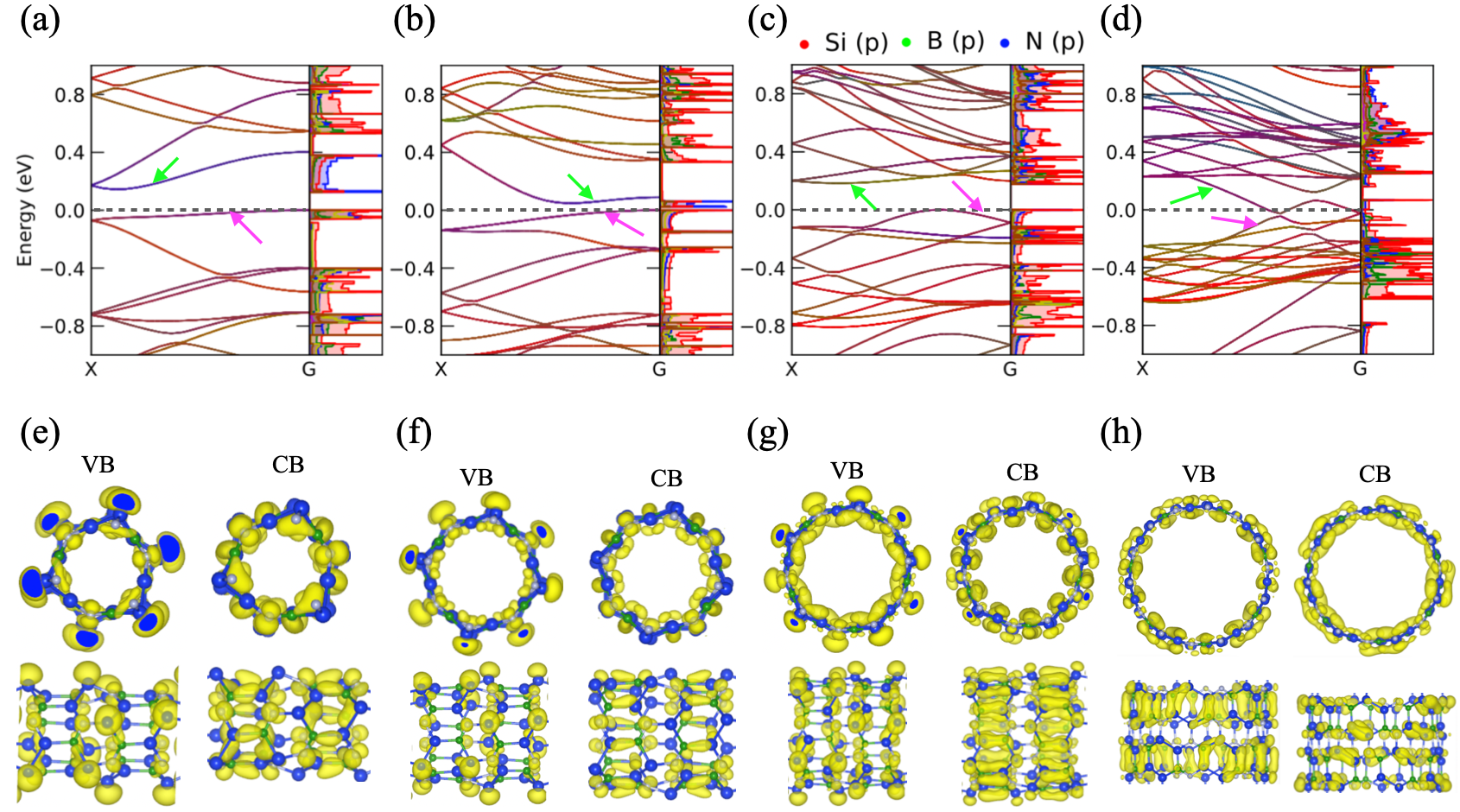}
	    \caption{Orbital- and element-resolved comparison of the electronic band structure and corresponding projected density of states for zigzag NTs (3,0), (4,0), (5,0) and (7,0), in separate columns. The set of top panels shows a traditional band-structure mapping while the set
	    of bottom panels provides analysis of the orbital nature that defines the
	    electrical transport; Nomenclature
	    as in corresponding analysis for the armchiar NTs, Fig.\ 6.
	    }
	    \label{band-zz}
\end{figure*}

The electronic properties of Si$_2$BN NTs are next discussed based on their curvature and the resulting hybridization effects \cite{blase1994hybridization}. 
We have tested for the impact of inclusion of the spin-orbit coupling and founded show any significant difference in the electronic band structure (see Fig. S8 in SM) for medium-size NTs. Our spin polarised calculations in suggest that there is no magnetism in these NTs. 

Fig. \ref{band-ac}(a-c) shows the projected electronic band structure and density of states for various armchair NTs. The lower panel in fig. \ref{band-ac} (d-f) stands for the band-decomposed density probability of the first band above and below the Fermi level. This represents the density for these band dispersion between G-X. 

In the upper panel of fig. Fig. \ref{band-ac}, armchair NTs show varying electronic properties such as semiconducting in (2,2), semi-metallic in (3,3), and metallic in large diameter (4,4) NTs. Table \ref{strain} also gives information about the chiral vector, diameter, and respective electronic properties. In (2,2) NT, the indirect bandgap was found to be 0.09 eV with a valence band (VB) maximum (VBM) in between X-G and conduction band (CB) minimum (CBM) at G-point. 

Fig. \ref{band-ac}(d) shows that the CB part that is located just above the Fermi level (shown by dotted line) arises primarily from the G-point and is defined by Si-B $\pi$ bonding. In contrast, the VB part that is located just below the Fermi level arises from contributions that are dominated by the $\pi$-$\sigma$ hybridization in Si-B atoms, while the density at N atom is partially localized. In the case of the (3,3) NT, bands touch the Fermi between G and X band points, reflecting a semi-metallic nature of conduction.  Band decomposed density in fig. \ref{band-ac}(e) shows that Si-Si and Si-B $\pi$ bond, which provide a downshift to the band above the Fermi. 
The electron/hole probability density is more localized in the Si-N bond region. In the top VB, there is still a $\pi$-$\sigma$ hybridization in the Si-B bond, which is confirmed by the $\pi$ orbital primarily localized at the outer surface of the tube at the Si atom.  A similar case was reported in the C NT \cite{blase1994hybridization}. 

For the relatively larger NT (4,4), the CB crosses the Fermi (making it metallic) and arises as dominated by Si-Si and Si-B $\pi$ bonds, similar to those found in the Si$_2$BN planar sheet. There is a change in band dispersion because of the larger diameter of the NT.  Here we find states at N atoms that are even more localized for the CB, Fig.\ 5(f). A similar behavior is evident in VB below the Fermi level, where a strong $\pi$ bond between two Silicon and one Boron atom is formed. 

We also extended our transport study to the  ultra-small (1,1) and large (7,7) armchair NTs (with diameters of $\sim$0.4 nm and $\sim$2.5 nm, respectively). It is important
to note that we find very small NTs to 
be unstable, see SM. Nevertheless, they are still worth investigating to fully
reveal the size impact on conduction trends.

We find that these (unstable, small) armchair NTs show a metallic nature of conduction. That is, a metal-type bandstructure is documented in Figs.\ S5 of the fig S6 SM. It arises because of the strong curvature effect and dominant $\pi$-$\sigma$ hybridization owing to structural deformations. These hybridization effects and stability are further supported by COHP in Fig.\ S7. A further discussion can be found in the SM.


Moving to zigzag NTs, Fig. \ref{band-zz} shows the band structures and band decomposed density for (3,0), (4,0), (5,0) and (7,0) NTs. Electronic band gap properties along with strain energy and diameter can be seen in table \ref{strain}. 

The small diameter tube (3,0) shows an indirect bandgap of 0.14 eV, where VBM is at G-point and CBM is at X-points. Fig. \ref{band-zz} (a and e) show that the highly localized $\pi$ orbital on N atom dominates the CB close to Fermi level and there is $\pi$-$\sigma$ hybridization Si-B. There is also a $\pi$-bond formation at an inner surface of the tube in between the second nearest Si-B atoms. The quasi flat VB component just below the Fermi level is dominated by localized states at alternate Si atoms, which has localization at the outer surface of the tube. The VB is partially constituted by the $\pi$-$\pi$ interaction in alternate Si and B atoms at the inner surface of the tube and localized states at N atoms. This 
dispersion-less feature comes from a highly distorted structure, where buckling is such that one Si atom in a hexagonal unit comes out of the plane resulting in the strongly localized states. We find that the strong localization is responsible for the bandgap opening.

Increasing the diameter, we next consider the(4,0) NT. Here, the lowest CB component is pushed down towards the Fermi level resulting in a reduced indirect bandgap of 0.05 eV. The VBM and CBM are at the G-point and in between the G and X points, respectively, in fig. \ref{band-zz} (b). It is clear from 
Fig.\ \ref{band-zz}(f) that the lowest CB is still dominated by the $\pi$ orbital localized in N atoms and Si-B $\pi$ bonds. Unlike (3,0) tube, this lacks the inner surface Si-B bonds between second nearest neighbors due to large diameter and shows less buckling effect. The quasi-flat band in VB has localized behavior out of the curvature Si and N atoms similar to (3,0) NT in 
Fig.\ \ref{band-zz}(e). 

For the (5,0) NT, Fig. \ref{band-zz}(c), we find that the CBM is located at X-point and the VBM is located in between the G-X with an indirect bandgap of 0.18 eV. Figure \ref{band-zz}(g) shows a strong $\pi$ bond behavior in Si-B bond at both the inner and outer surface of the tube and localized state at the out of surface Si atom in the lowest CB. However, there is less buckling in comparison to (3,0) and (4,0) NTs. There is a minimal contribution from the N atom as can also be seen in fig. \ref{band-zz} (c). The VB close to the Fermi level mainly arises by the localized state on the Si atom at the outer surface and Si-Si $\pi$ bonds in the inner surface of the tube. There are also contributions from B-N $\pi$ bonds. 

Figure \ref{band-zz}(d) shows that the CBM and VBM come close to each and form cone-like behavior in the $\sim$1.5 nm diameter NT (7,0). Band decomposed density picture in Fig. \ref{band-zz} (g) shows that Si-Si and B-N $\pi$ bond nature is restored in the conduction band. The VB arise from Si-B $\pi$ bonds, which is similar to those of the Si$_2$BN sheet. The highly delocalized $\pi$ bond behavior is responsible for a metallic nature in the (7,0) NT. A similar electronic structure was also found in (6,0) and even larger diameter tubes (12,0). The smallest possible diameter zigzag NT (2,0) also shows similar semiconducting behavior due to the strongly localized state around the Fermi level as described in SM.

The effect of even larger diameters is also investigated for the  $\sim$2.5 nm diameter armchair and zigzag NTs. Similar to the Si$_2$BN sheet, these NTs remain metallic with more states around the Fermi level and can be seen in Fig.\ S9 and fig S10. We  find  band-projected densities and a bonding nature that are similar to those found in a planar Si$_2$BN sheet.

\begin{figure}[h!]
	    \centering
	    \includegraphics[width=1\linewidth]{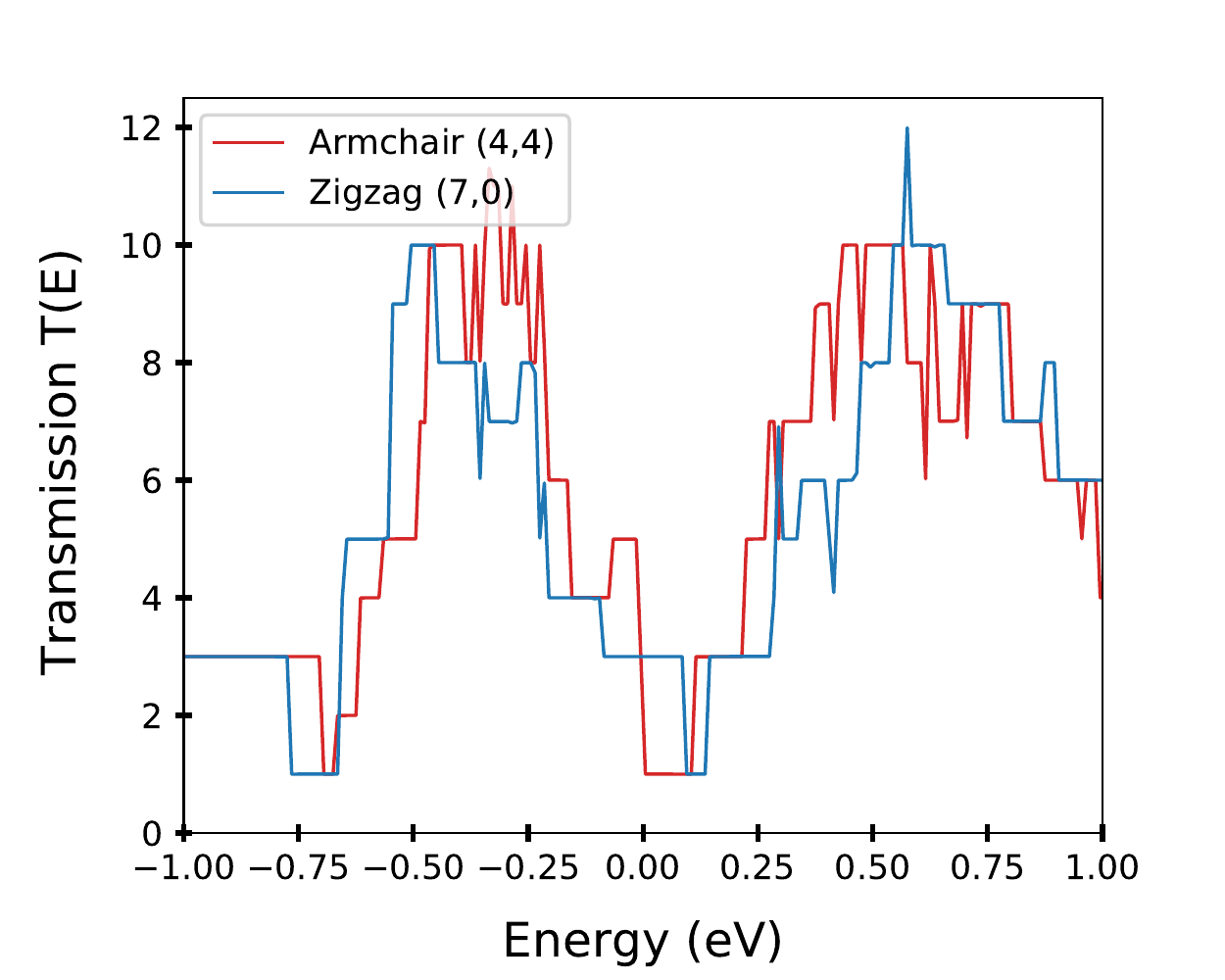}
	    \caption{Electron transport behaviour at zero-bias transmission in the armchair Si$_{2}$BN NT (4,4)
	    (green color) and the zigzag (7,0) NT (red color). We find a step wise transmission that can be correlated with the associated bands in electronic band structure these NTs, see tet. Both NT have essentially the same diameter $\sim$1.5 nm.}
	    \label{trans}
    \end{figure}  

Interestingly, the small-diameter NTs are found to be semiconducting. This behavior is different than what is found in carbon NTs, BC$_2$N, and Silicene NTs \cite{blase1994hybridization,azevedo2006stability,liu2021essential}. Hybridization is the key to understanding the metallic behavior in carbon NTs at small diameters \cite{blase1994hybridization}. We find a different Si$_2$BN behavior (in the small-diameter limit) because the constituent elements have varying electronegativity values.

At a small diameter, the Si$_2$BN hexagonal ring gets further distorted and buckles out of the plane with stretched bonds. This gives density localization despite the strong possibility of hybridization due to curvature. Similar behavior was reported by Shukla et al. \cite{shukla2017curious}, where puckered Si$_2$BN sheet showed semiconducting behavior. In the large-diameter tubes, bond angles further deviate from the ideal 120$^\circ$ on the NT wall, this results in an increase of the $\pi$-$\sigma$ orbital interaction in comparison to the planer sheet and alter the reactivity of the NT walls with bonding strengths varying with the binding sites. Our Bader charge analysis for armchair (4,0) nanotube suggests that the Si sites are expected to be more reactive (sticky) compared to other sites. As such, the new NTs have a different behavior than the distinct chemistry of C NTs and hBN NTs.

To further evaluate the electrical conductivity we also provide a set of nonequilibrium Green function (NEGF) transport calculations \cite{LippSchwing,KadBaym,Keldysh,LangrethPhase,Baranger89,KondoPRL,YigalNed,WiNobelSymp,HyHeDa94,Lang95,DiVentra00,DiVentra01,Stefanucci,LSCDFT,LSCDFTforces}. There exist a rigorous 
Lippmann-Schwinger (LS) scattering formalism \cite{LippSchwing,LangrethPhase,LSCDFT,LSCDFTforces} for such studies.
Even if we use exchange-correlation functionals crafted for ground-state DFT \cite{Verdozzi06}, our computation of effective single-particle LS orbitals have a well-defined physics meaning in terms of Friedel scattering phase shifts and grand-canonical
ensemble DFT \cite{Mermin65,HarrisFriedelA,HarrisFriedelB,LSCDFTforces}. 
In the ballistic transport case, electron-phonon and phonon-phonon scattering  is ignored and computation of the electron transmission $T(E)$ essentially reduces to a counting of the number of available conduction channels, each
having some probability for being affected by elastic scattering. Each channel carries a so-called conduction quanta, $G_0 = 2e^2/h$, where $e$ is electron charge and $h$ is plank's constant. The resulting electron conduction can thus be asserted from $G = G_0 T(E)$. This simple ballistic-transport characterization \cite{Lang95,DiVentra00,TranSiesta02,Verdozzi06} that can be used for a first assessment of the Si$_2$BN NT electron-transport performance.

Figure \ref{trans} reports our electron transport study of the armchair (4,4) and zigzag (7,0) NTs, both having a $\sim$1.5 nm diameter. We find that the computed transmission function $T(E)$ has essentially regular steps, reflecting that elastic scattering does not strongly impede the electrons dynamics across central Si$_2$BN model region in our NEGF calculation setup, See Figure S11 in SM.

Fig. \ref{trans} shows that the transmission is lower at the Fermi level for the armchair case than for the zigzag case. This is also expected from  from the band structure results shown in Figs. \ref{band-ac} and \ref{band-zz}. The
zigzag (armchair) NT has three (one) transmission channel at the Fermi level. However, the we find that applying a bias can significantly increase the armchair NT transmission. In contrast, the zigzag NT behavior is a flat transmission line above the Fermi level 0 eV to 0.1 eV. This is due to a cone-type band dispersion around the Fermi level.

Finally, we consider the electronic part of thermal conductivity. In metallic systems the contribution is directly proportional to the electrical conductivity, assuming that the Wiedemann-Franz Law \cite{minnich2009bulk}
remains valid also for our low-dimensional focus. We therefore expects that 
also the  electronic part of thermal conductivity will be high in Si$_2$BN nanotubes \cite{t2019thermoelectric}. 




\section{Summary and outlook}
\label{conclusion}

We have systematically investigated the structural, stability, chemical bonding of the armchair and zigzag Si$_2$BN nanotube using ab-initio simulations as well as analys
of the bonding nature. 
The resemblances of the Si$_2$BN and carbon NT formation suggest a path for controlling the electronic structure. The presence of distorted hexagonal patterns and of covalent-radii differences suggest an even richer behavior.

Structural stability is confirmed by the cohesive and strain energies, phonon dispersion spectra, and first-principles molecular dynamics (FPMD) at 1000 K. The set of stability analyses suggests that an experimental realization of Si$_2$BN NT is possible, and might be realized similarly to what has been demonstrated for BC$_2$N. Our COHP analysis confirms that strong covalent bonding between the Si-Si, Sb-B/N, and B-N atoms, persist also after the nanotube formation (across a range of nanotube sizes). This suggests that the NT construction is also practically feasible, in that, the bonds can withstand the large additional strain that such processes entail.

We also find that the predicted electronic properties show significant variation. In effect, there are possibilities for controlling the resulting electronic structure by varying the NT chiral vector and diameters. We predict a semiconducting behavior at small diameters and metallic and semi-metallic natures at larger diameter. 

Finally, we expect the implications to impact the NT electrical and thermal conductivity, stability, and surface reactivity. As such, we expects that Si$_2$BN NTs may find potential applications in nanoscale devices, batteries, sensing, hydrogen storage, and photocatalysis.

\begin{acknowledgements} 
D.S., N.K. and R.A. acknowledge Swedish Research Council (VR-2016-06014 $\&$ VR-2020-04410) and J. Gust. Richert stiftelse, Sweden (2021-00665) for financial support. V.S. and P.H.  acknowledge the Swedish Foundation for Strategic Research (SSF) through grant IMF17-0324, as well as the Chalmers Area-of-Advance-Materials theory \& -Production theory activities. The authors gratefully acknowledge computational resources from the Swedish National Infrastructure for Computing SNIC (2021/1-42 as well as SNIC2020-3-13, SNIC2021-3-18), HPC2N and C3SE.
\end{acknowledgements} 
\bibliography{Ref}

\cleardoublepage
\pagebreak
\widetext
\begin{center}
\textbf{\large Supplementary Materials for:\\
Stability of and conduction in single-walled Si$_{2}$BN nanotubes}
\end{center}
\setcounter{equation}{0}
\setcounter{section}{0}
\setcounter{subsection}{0}
\setcounter{figure}{0}
\setcounter{table}{0}
\setcounter{page}{1}
\makeatletter
\renewcommand{\theequation}{S \arabic{equation}}
\renewcommand{\thetable}{S \Roman{table}}
\renewcommand{\thefigure}{S \arabic{figure}}
\renewcommand{\bibnumfmt}[1]{[S#1]}

\section{Electronic properties of planar S{i}$_{2}$BN monolayer}
    \begin{figure}[htp!]
	    \centering
	    \includegraphics[width=1.0\linewidth]{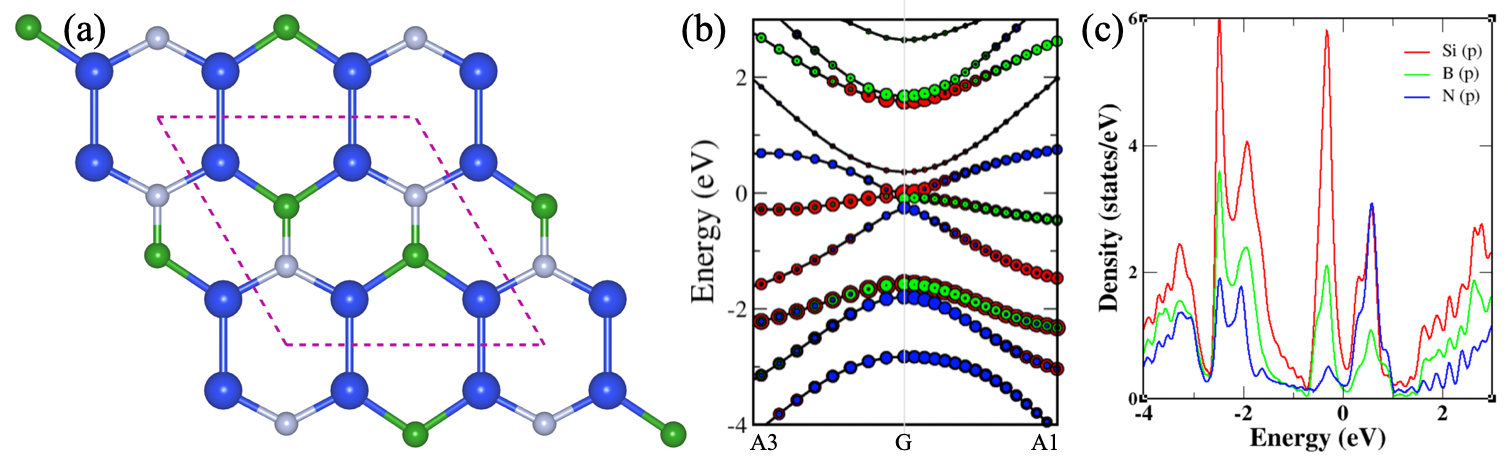}
	    \caption{(a) Fully relaxed two dimensional Si$_2$BN structure, (b) orbital contributed electronic band structure and (c) projected density of states of graphene-like Si$_{2}$BN monolayer sheet.} 
	    \label{2D-SBN}
    \end{figure} 
    
\section{Structural stability}
    \begin{figure}[htp!]
	    \centering
	  \subfloat{{\includegraphics[width=0.4\linewidth]{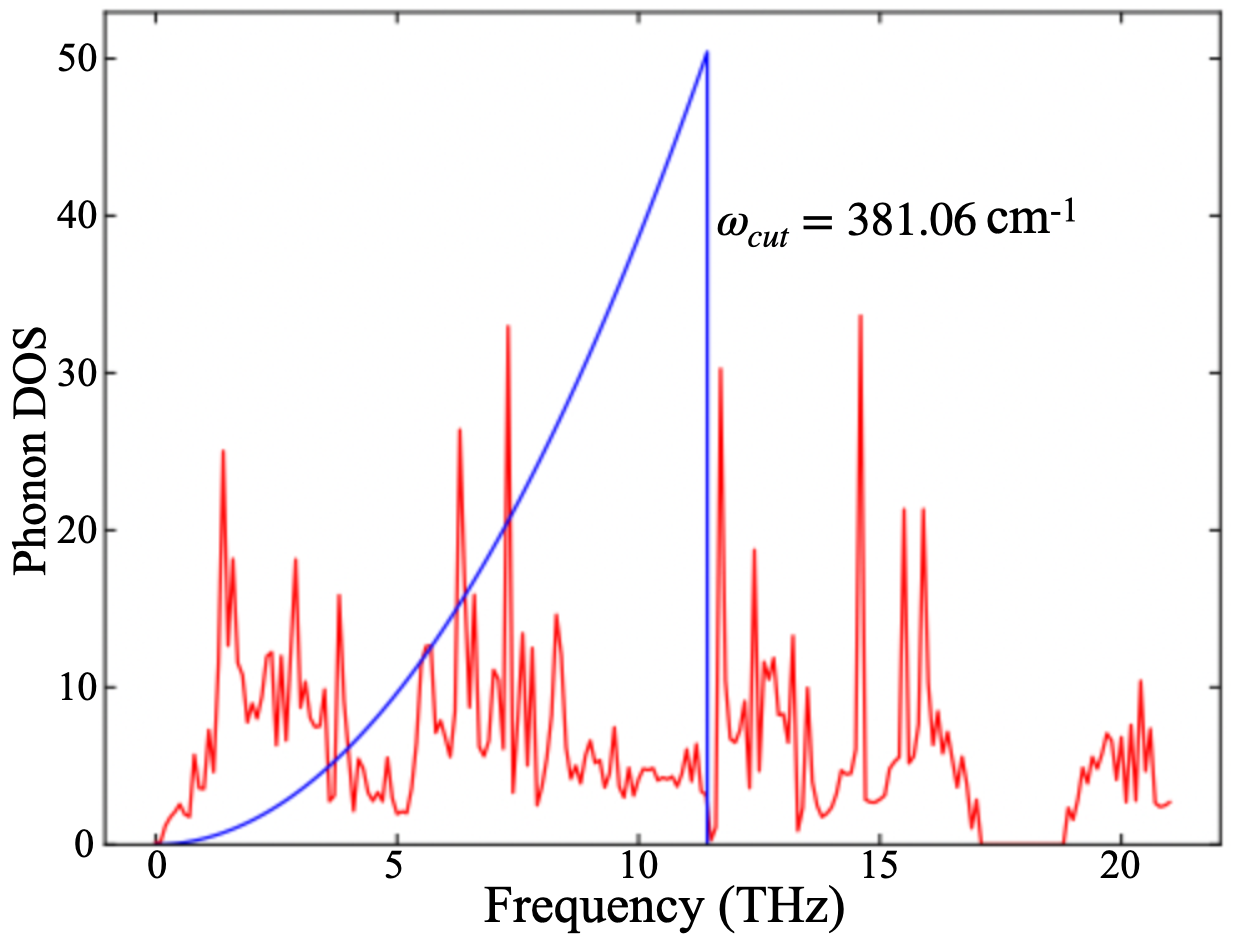}}}%
        \qquad
    \subfloat{{\includegraphics[width=0.42\linewidth]{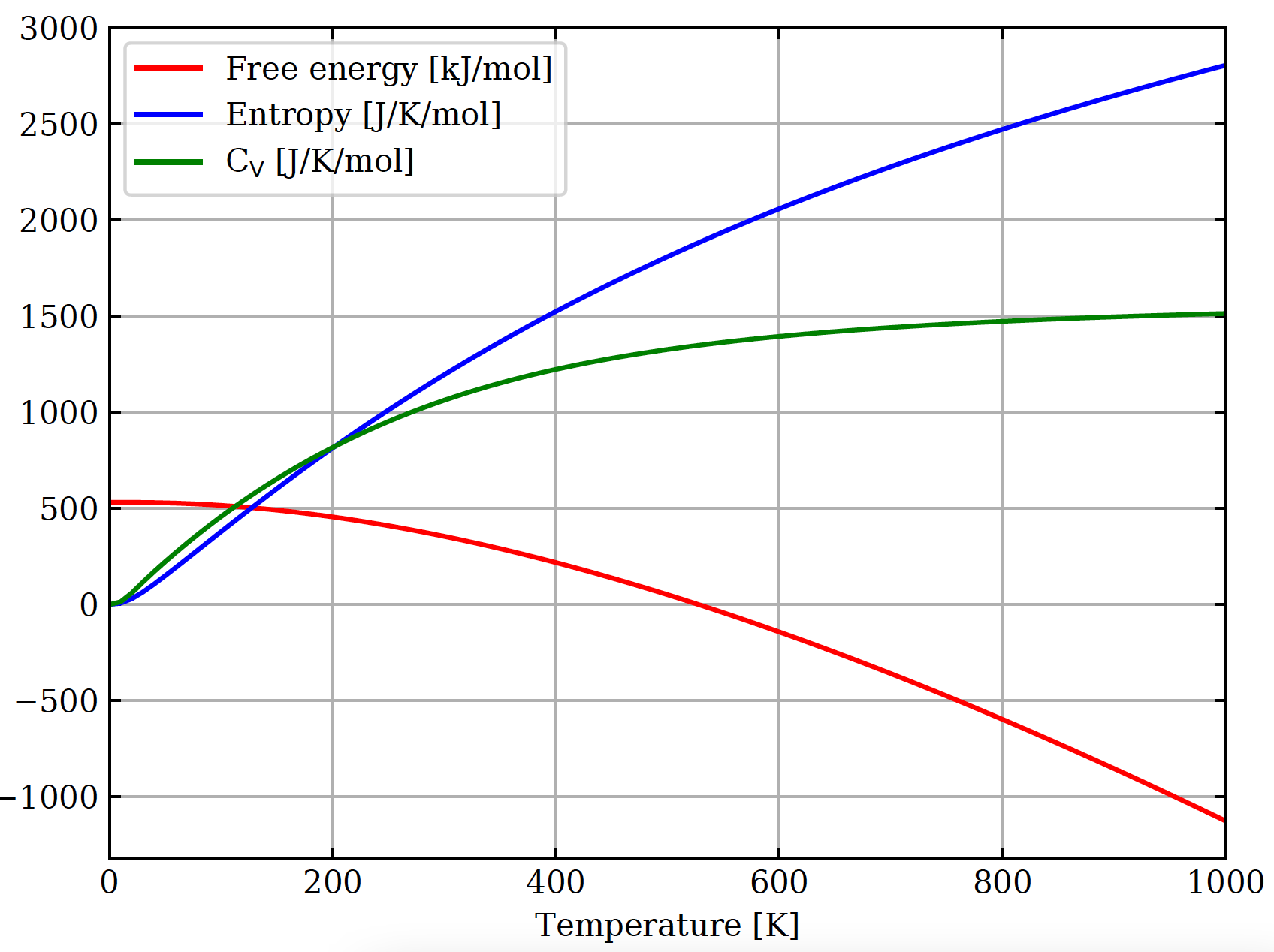}}}
	    \caption{Phonon vibrational density of states of armchair Si$_{2}$BN (4,4) NT. Upper panel shows the cutoff frequency. The lower group of phonons can be seen as connected to the acoustic mode and we shall refer to $\omega_{cut}$  as an effective acoustic cutoff described with blue line. Lower panel shows the Free energy, entropy, and specific heat varying with temperature.} 
	    \label{Phon-dos}
    \end{figure}

    \begin{figure}[hb!]
	    \centering
	    \includegraphics[width=0.7\linewidth]{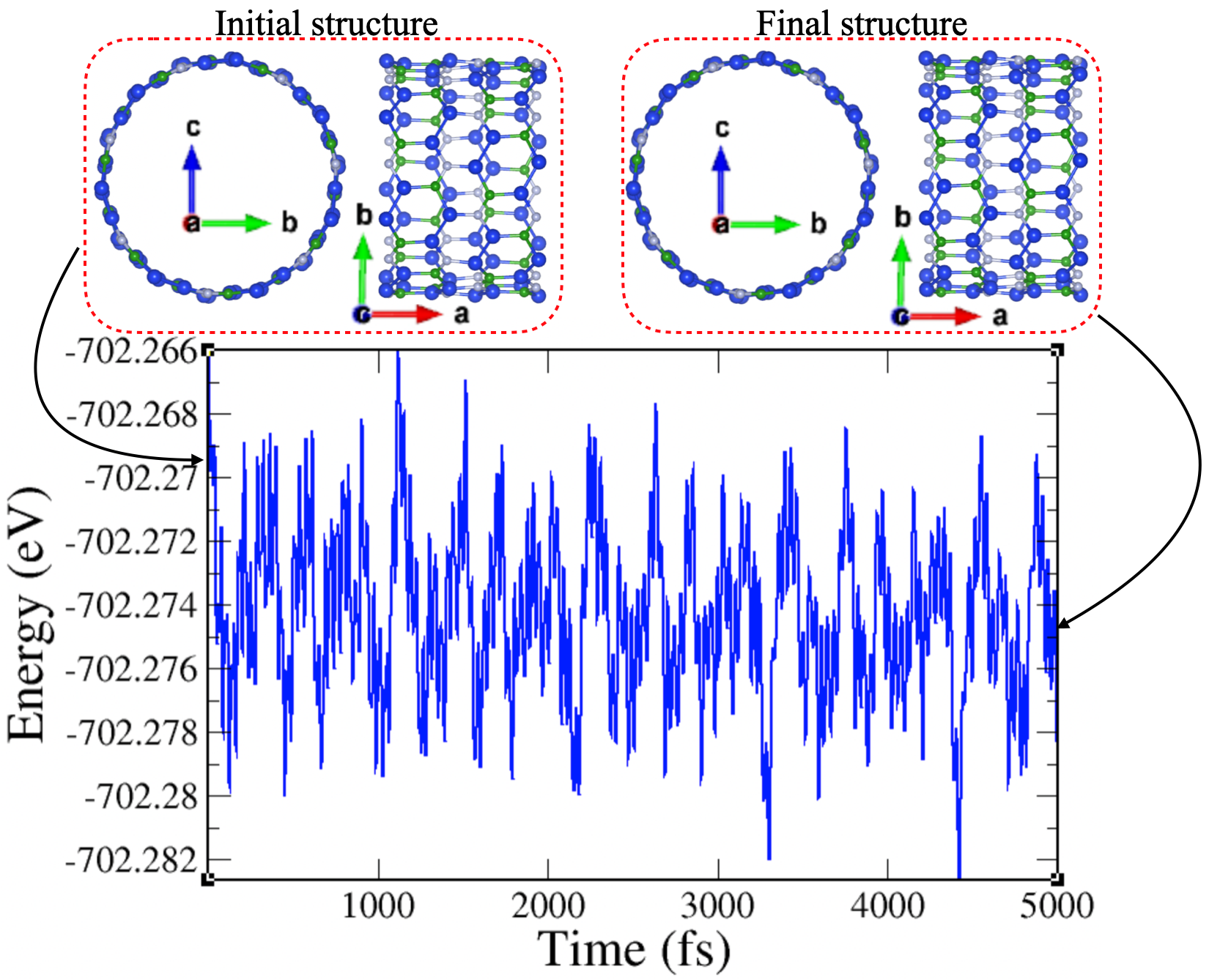}
	    \caption{Variations of the total energy for zigzag (7,0) Si$_{2}$BN nanotube during first principle molecular dynamics simulation at higher temperature 1000 K for 5 ps. The variation in total energies are extremely small which shows that this nanotubes is highly stable at relatively higher temperatures.The inserts above the energy variations shows the initial and final structure during the molecular dynamics simulations with similar bond length and structure.} 
	    \label{MD-zigzag}
    \end{figure}

\section{Smallest diameter nanotubes}
The smallest possible rolling of Si$_2$BN sheet is (1,1) nanotube in armchair direction and (2,0) nanotube in the zigzag direction in fig. \ref{small-nt}. These nanotubes get heavily distorted after relaxation and break the hexagonal Si$_2$BN ring. Phonon density of states clearly shows soft phonon in the armchair direction where distortion is more and in principle there is nanotube structure remains. High-temperature FPMD simulations also show large variations in energy. On the other hand, the smallest zigzag nanotube is possible with (2,0) chiral vectors. Unlike armchair, the smallest zigzag nanotube does not break the hexagonal pattern of Si$_2$BN but there is strong buckling out the smooth curvature for on Si atom in Si$_2$BN hexagonal ring. It still keeps the nanotube kind of structure. This is further confirmed by the phonon structure band where we see there is no soft phonon. FPMD simulation also reveals that there is not so much thermal variation during the run.

The smallest armchair nanotube has evident soft phonons around the G point and along the band line between G-X. The armchair nanotube gives stretched bonds during relaxation, which clearly stands with a less stable structure. In the zigzag case, there are no imaginary frequencies in the phonon band structure and relatively bonds are less stretched during the full relaxation. This indicates a stable structure. 

    \begin{figure}[htp!]
	    \centering
	    \includegraphics[width=1.0\linewidth]{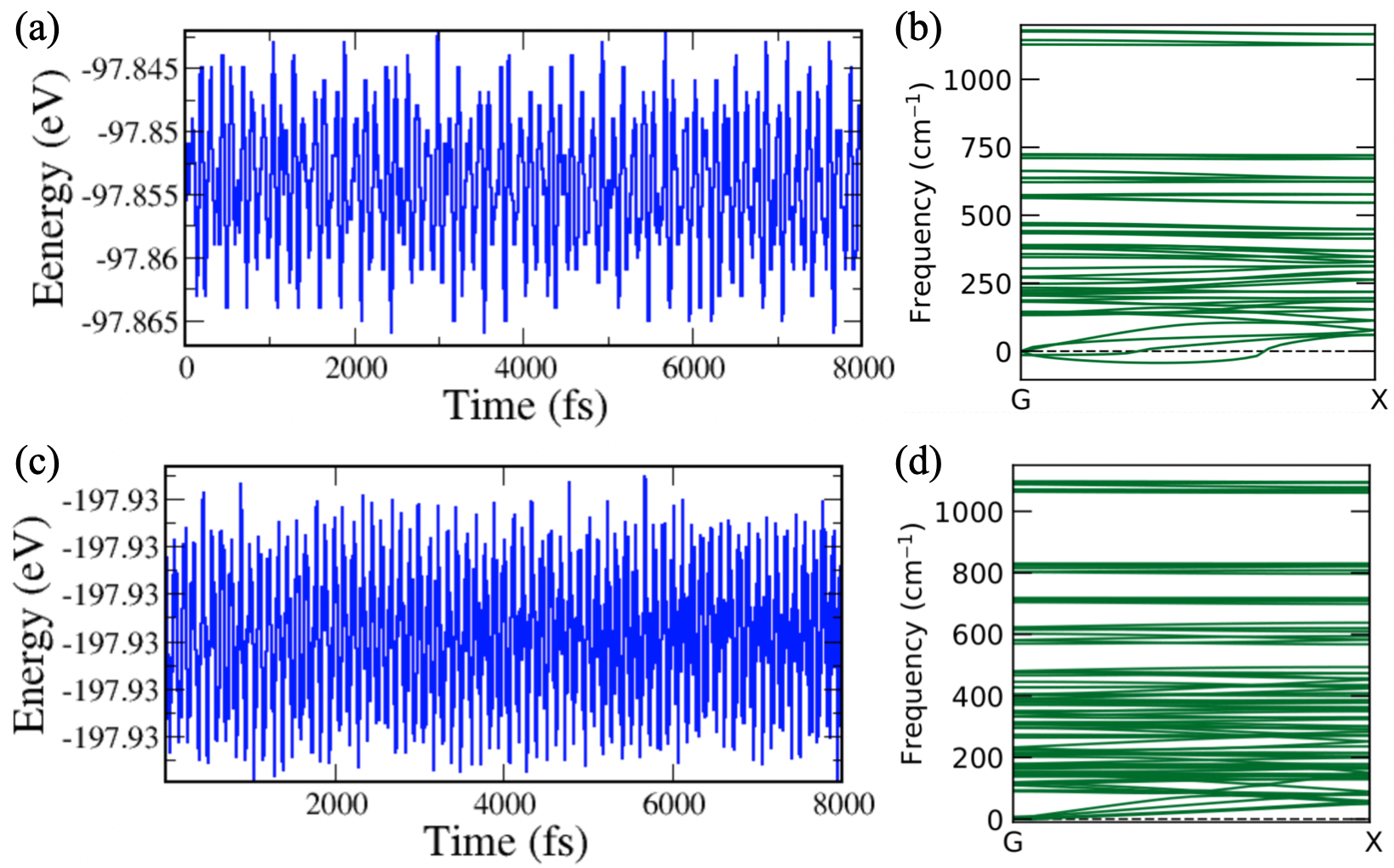}
	    \caption{Ab-initio molecular dynamics (AIMD) simulations of smallest Si$_{2}$BN NT along (a) armchair (1,1) and (c) zigzag (2,0) direction. Phonon dispersion spectra of smallest Si$_{2}$BN NT along (b) armchair (1,1) and and (d) zigzag (2,0) direction.} 
	    \label{aniso-sm}
    \end{figure} 

\section{Electronic properties}

\begin{figure}[htp!]
	    \centering
	    \includegraphics[width=0.5\linewidth]{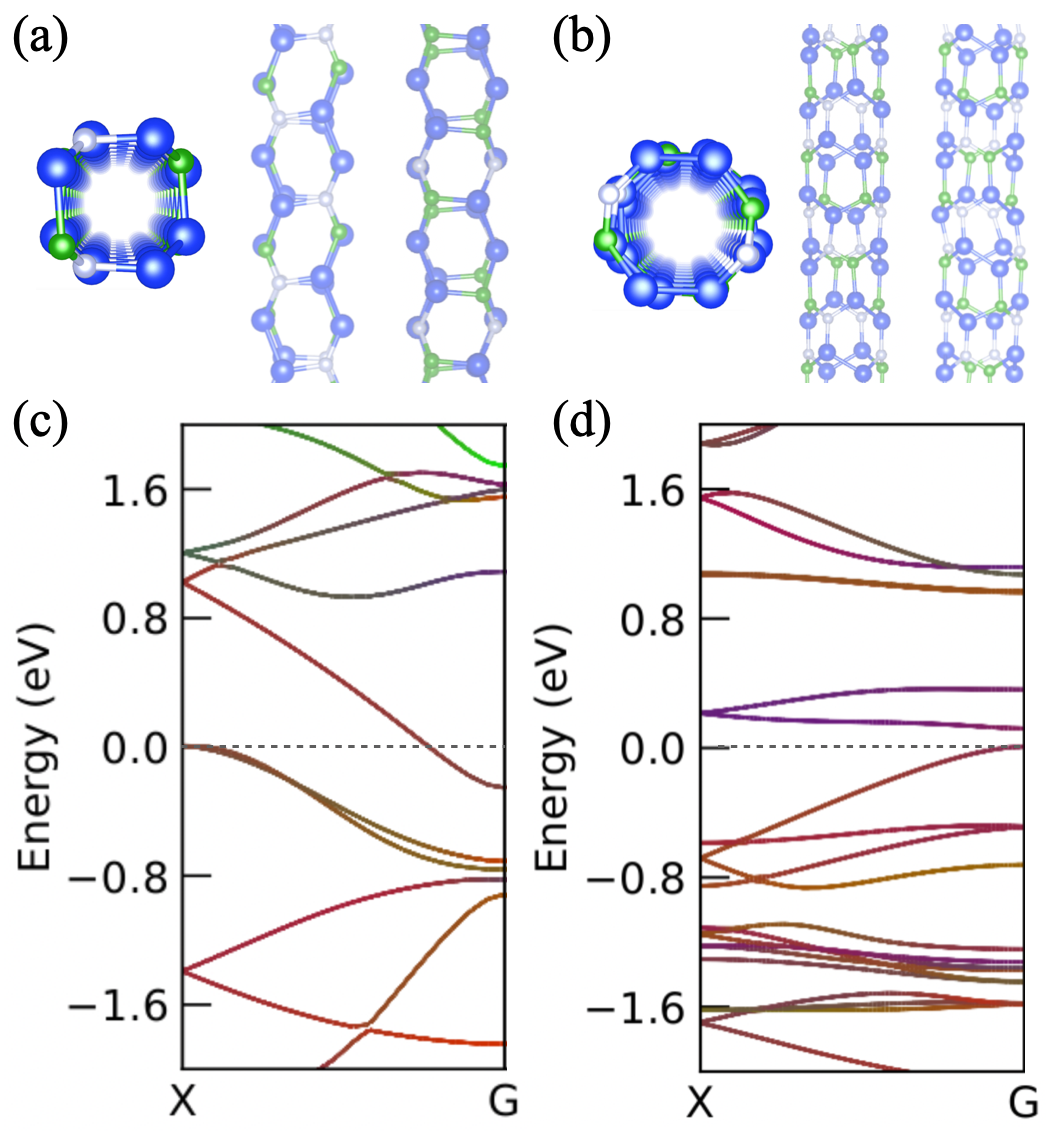}
	    \caption{Schematic illustration of smallest configuration nanotubes (a) (1,1) armchair ,(b) (2,0) zizgag direction. Electronic band structure of these smallest configuration nanotubes, (c) (1,1) in armchair and (d) (2,0) zigzag direction.} 
	    \label{small-nt}
    \end{figure}

\begin{figure}[htp!]
	    \centering
	    \includegraphics[width=0.7\linewidth]{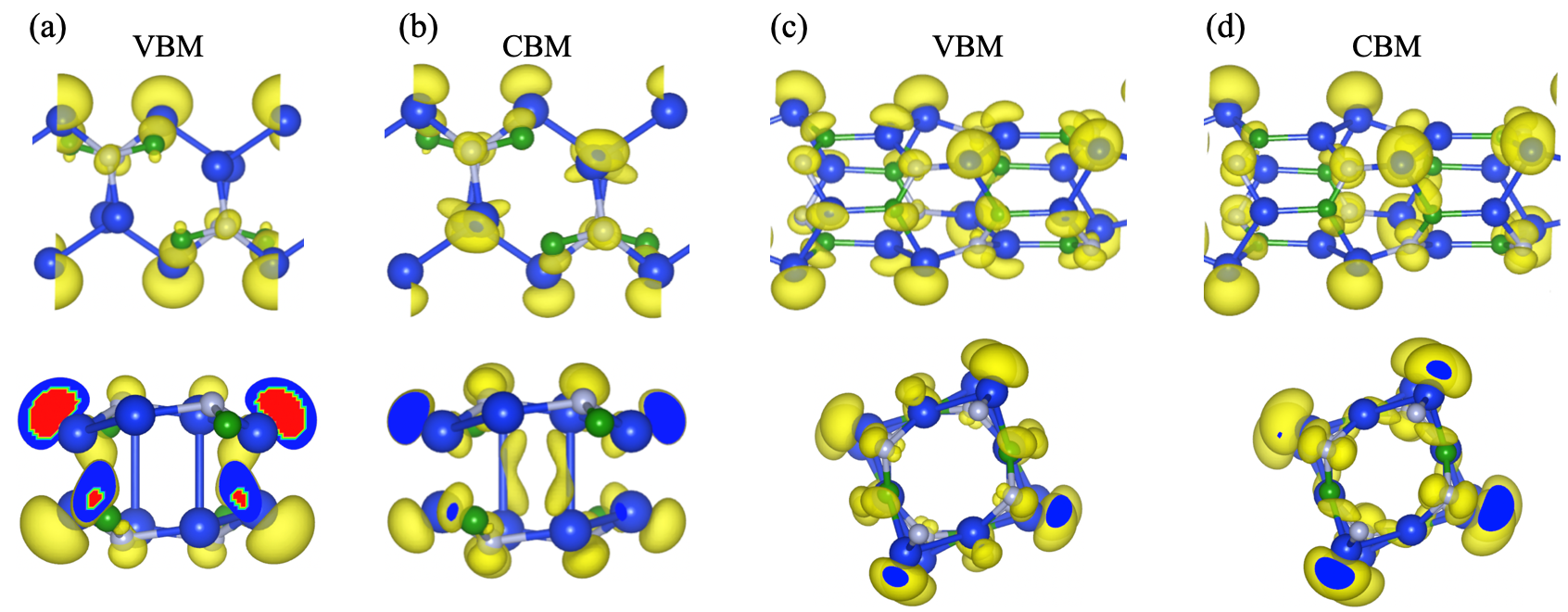}
	    \caption{The structural configuration of after full relaxation and band composed density at states close to the Fermi level in valence and conduction band. (a,b) stands for armchair (1,1) abd (c,d) for zigzag (2,0) Si$_{2}$BN NT with $\sim$0.4 nm.} 
	    \label{vbm-cbm-sm}
    \end{figure}

We discussed that armchair nanotube breaks in the smallest diameter configuration but still, we tried to calculate the electronic properties. Fig. \ref{small-nt} shows the band structure for the smallest nanotube possible. Fig \ref{small-nt}(c) shows the metallic properties because the hexagonal rings are broken. There is a strong $\pi$ bond between in Si-Si and further stronger $\pi$-$\sigma$ hybridization. The nature of hybridization is evident in fig. \ref{vbm-cbm-sm} (a,b). On the other hand, zigzag (2,0) nanotubes follow the trend of (3,0), (4,0) in the main paper. There is a direct bandgap of 0.11 eV that can be visible in fig. \ref{small-nt} (d). Band decomposed density in fig. \ref{vbm-cbm-sm} (c) shows that the band above the Fermi level is nearly flat because of highly localized states on nitrogen. The band below the Fermi is constituted by the Si-B $\pi$ bond, and localized state in out-of-the plane silicon atoms (see in fig. \ref{vbm-cbm-sm} (c,d)). Despite strong curvature, this gap is opened because of highly localized density and stretched bonds. 

\begin{figure}[htp!]
	    \centering
	    \includegraphics[width=0.7\linewidth]{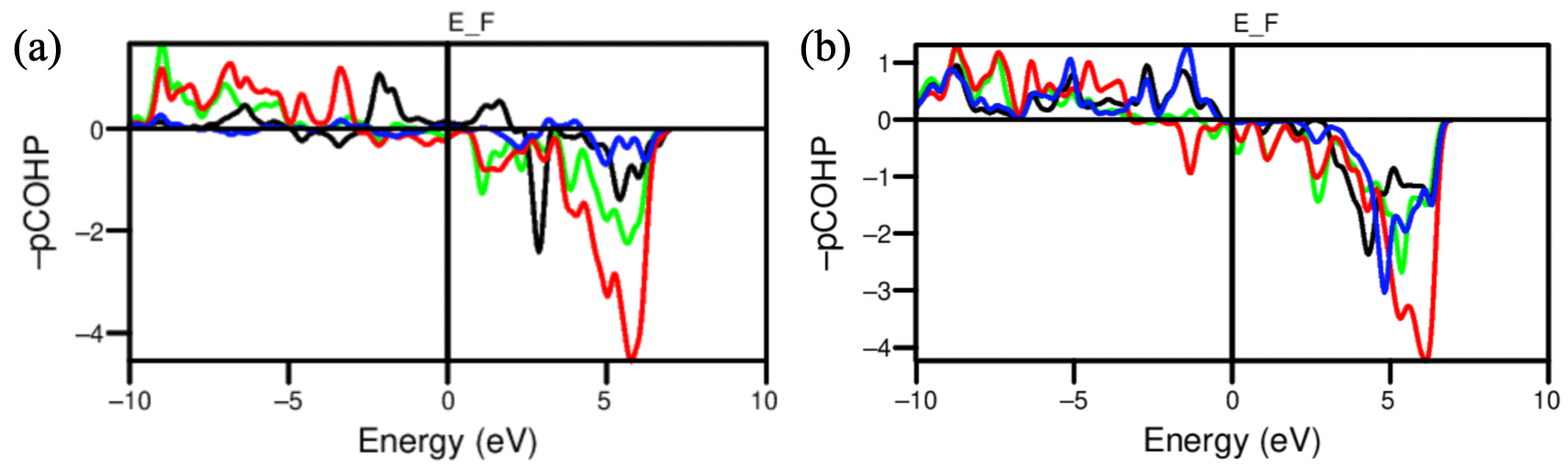}
	    \caption{Projected crystal orbital Hamilton Population (-pCOHP) analysis of smallest Si$_{2}$BN NT for (a) armchair (1,1) and (b) zigzag (2,0) direction summed up over the pairs Si-Si, B-N, Si-B and Si-N. The zero level shows the Fermi energy ($E_{F}$). The -pCOHP provides information about the contribution of a specific bond (bonding or anti bonding) to the band energy. The positive value on y-axis represent the bonding contribution and negative value on y-axis shows the antibonding contribution. The black, red, green and blue color represents the Si-Si, B-N, Si-N and Si-B pairs, respectively.} 
	    \label{bond-analysis-sm}
    \end{figure}

To further investigate the chemical bond stability we tested plane-wave base crystal orbital Hamilton population (pCOHP) analysis for the smallest armchair and zigzag of Si$_2$BN nanotubes respectively. We can clearly see in fig. \ref{bond-analysis-sm} (a), the anti-bonding states below the Fermi level in armchair for Si-Si, Si-B bonds. there are also some bonding states above the Fermi level that come out for Si-Si bonds gives information about the conducting behavior. Moving to the zigzag direction there is antibonding behavior below the Fermi level for Si-B bonds and the rest of the bonds show the bonding nature similar to (7,0) zigzag nanotubes. There is no bonding state above the Fermi which goes well with the semi-conducting nature of the zigzag nanotube. 

The ICOHP values of Si-N, Si-Si, B-N, and Si-B atom pairs are found to be -5.34, -2.16, 9.20, and -1.02 eV for armchair (1,1) smallest Si$_{2}$BN NT, respectively. While the ICOHP values of zigzag (2,0) smallest Si$_{2}$BN NT for Si-N, Si-Si, B-N, and Si-B atom pairs are -5.90, -4.57, -8.73, and -4.77 eV, respectively. The negative values of ICOHP values between atom pairs show strong covalent bond interactions.

    \begin{figure}[htp!]
	    \centering
	    \includegraphics[width=0.7\linewidth]{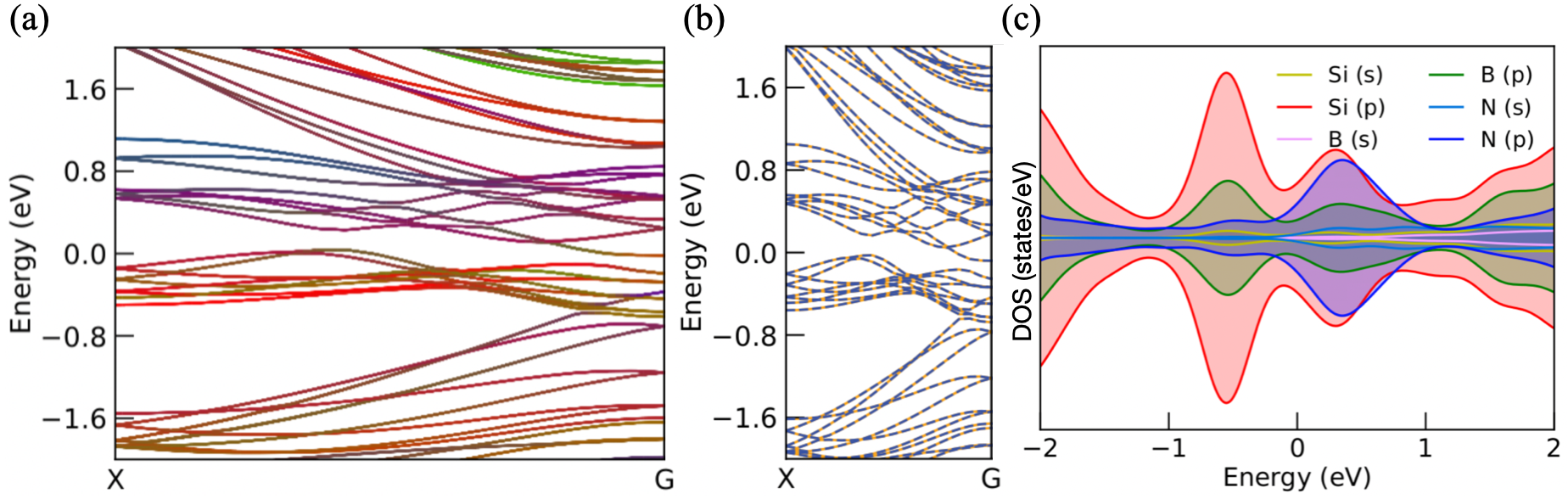}
	    \caption{(a) Orbital contributed electronic band structure with the effect of spin-orbit coupling for armchair Si$_{2}$BN (4,4) NT. The red, green and blue color represents the p-orbitals of Si, B and N atoms. Spin-polarized (b) electronic band structure and (c) projected density of states of armchair Si$_{2}$BN (4,4) NT using GGA-PBE functional.} 
	    \label{SOC}
    \end{figure} 
    
     \begin{figure}[htp!]
	    \centering
	    \includegraphics[width=0.7\linewidth]{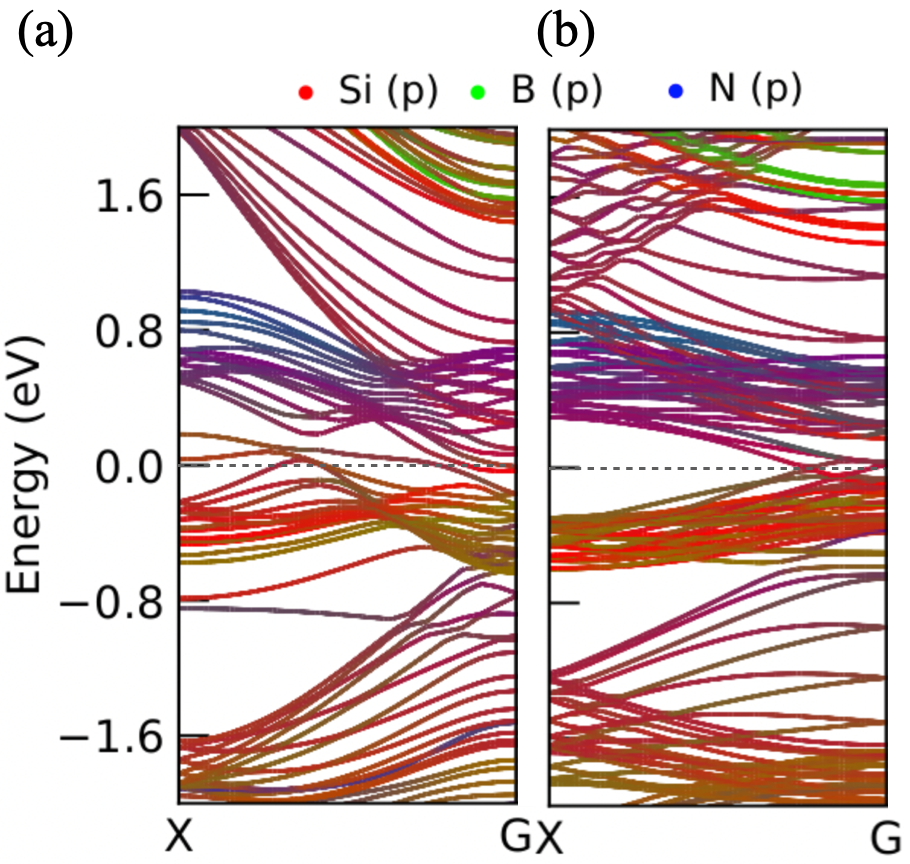}
	    \caption{Orbital contributed electronic band structures of (a) armchair (7,7) and (b) zigzag (12,0) directions of Si$_{2}$BN NT. The diameter of the Si$_{2}$BN NT are considered as $\sim$2.5 nm.} 
	    \label{band-lm}
    \end{figure}
\section{Large diameter nanotube}
\begin{figure}[htp!]
	    \centering
	    \includegraphics[width=0.8\linewidth]{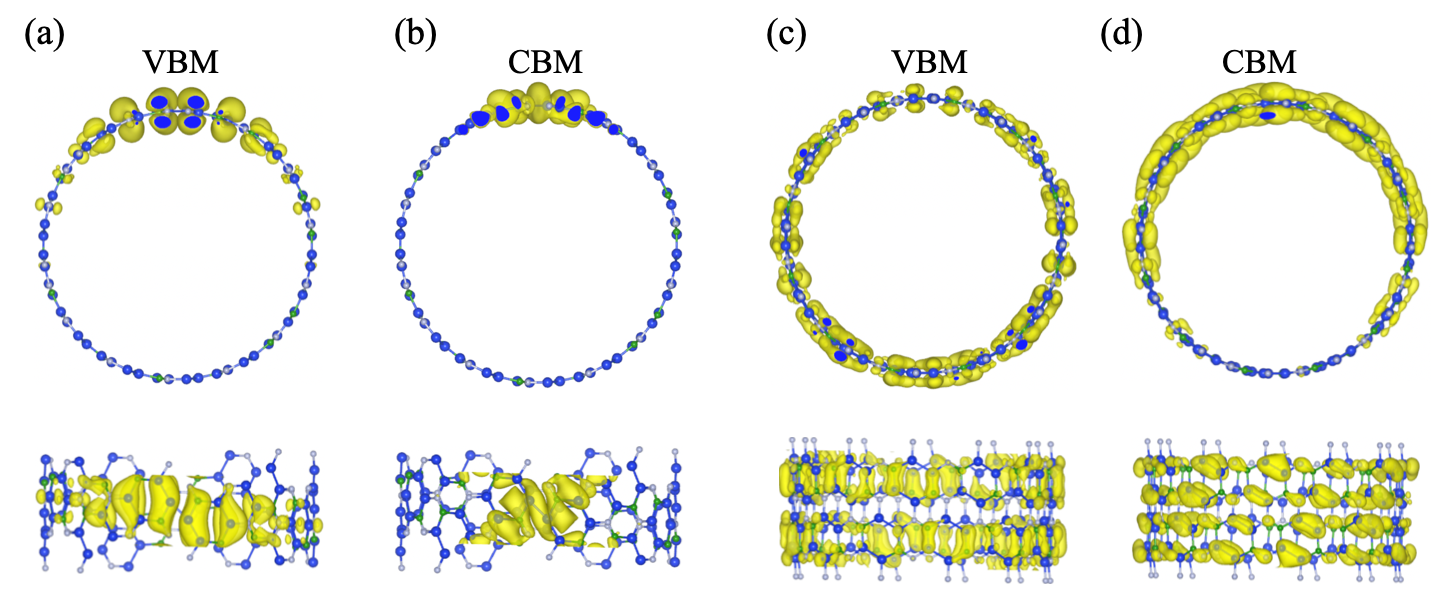}
	    \caption{Isosurfaces of band-decomposed charge density of the VBM, and CBM of (a,b) armchair (7,7) and (c,d) zigzag (12,0) directions of Si$_{2}$BN NT, respectively.} 
	    \label{vbm-cbm-lm}
    \end{figure}

Figure \ref{band-lm} shows the electronic band structure for the (7,7) armchair direction and (12,0) zigzag direction nanotubes with a large diameter of 2.5 nm. It is pretty evident that the electronic structures are similar to 1.5 nm nanotubes in the main paper but with more number bands around the Fermi level. Fig. \ref{vbm-cbm-lm} shows the bands around the Fermi level which shows similar behaviour like 1.5 nm tubes. 

     \begin{figure}[htp!]
	    \centering
	    \includegraphics[width=0.7\linewidth]{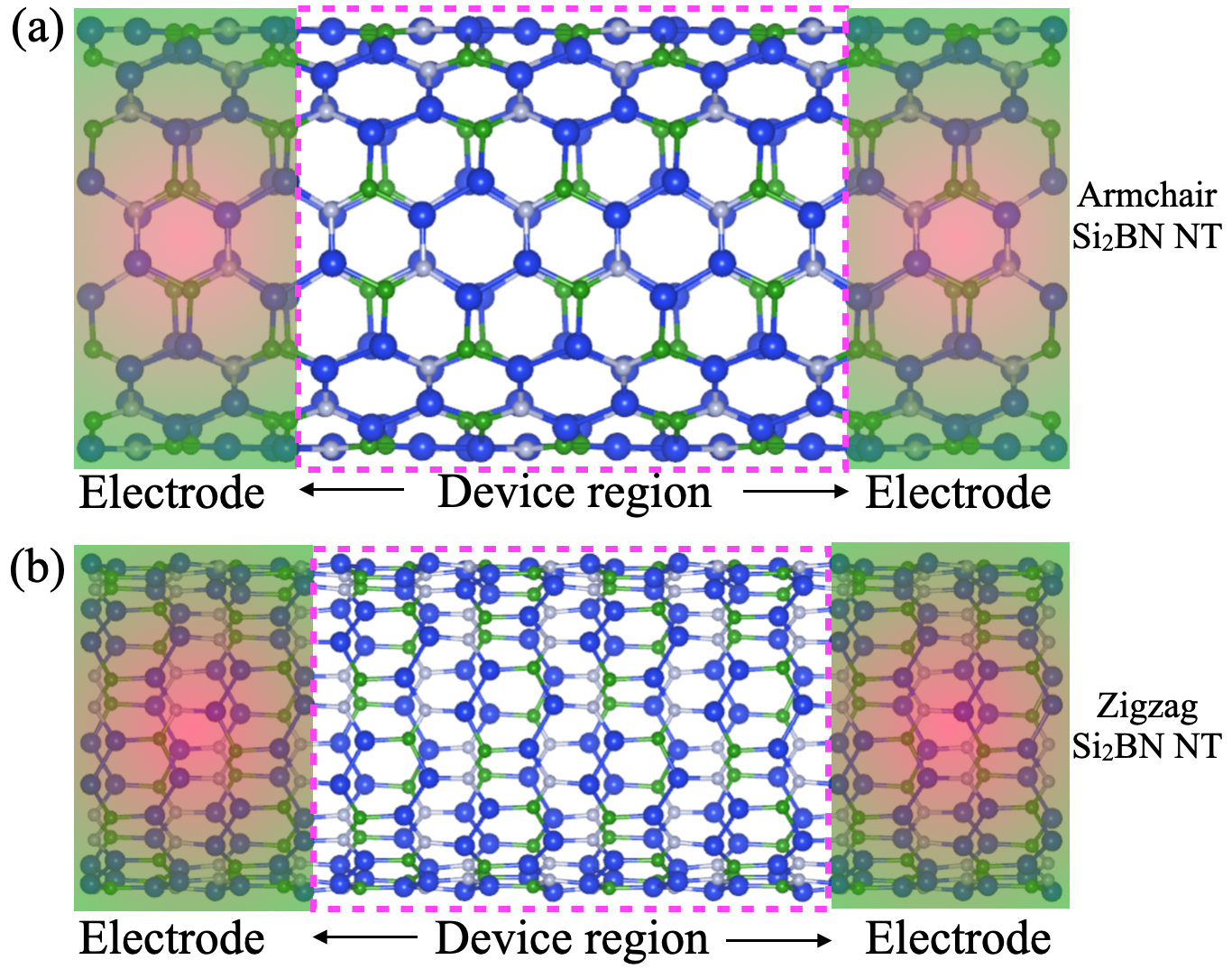}
	    \caption{Schematic diagram for NEGF calculations setup for (a) armchair (4,4) and (b) zigzag (7,0) directions of Si$_{2}$BN NT.  The central device regions are connected with semiinfinite electrodes represented as shaded region.} 
	    \label{Trans_setup}
    \end{figure}

\end{document}